\documentclass[preprint,showpacs,preprintnumbers,superscriptaddress,amsmath,amssymb]{revtex4}

\usepackage{graphicx}
\usepackage{dcolumn}
\usepackage{bm}

\begin{document}

\title{ 
Evidence of the ${\boldmath \Theta^+}$ 
in the ${\boldmath \gamma d \to K^+K^-pn}$ reaction}

\author{T.~Nakano}
\affiliation{Research Center for Nuclear Physics, Osaka University, Ibaraki~567-0047, Japan} 
\author{N.~Muramatsu}
\affiliation{Research Center for Nuclear Physics, Osaka University, Ibaraki~567-0047, Japan} 
\author{D.S.~Ahn}
\affiliation{Research Center for Nuclear Physics, Osaka University, Ibaraki~567-0047, Japan} 
\author{J.K.~Ahn}
\affiliation{Department of Physics, Pusan National University, Busan~609-735, Korea}
\author{H.~Akimune}
\affiliation{Department of Physics, Konan University, Kobe~658-8501, Japan}
\author{Y.~Asano}
\affiliation{Japan Synchrotron Radiation Research Institute, Mikazuki~679-5198, Japan}
\author{W.C.~Chang}
\affiliation{Institute of Physics, Academia Sinica, Taipei~11529, Taiwan}
\author{S.~Dat\'{e}}
\affiliation{Japan Synchrotron Radiation Research Institute, Mikazuki~679-5198, Japan} 
\author{H.~Ejiri}
\affiliation{Japan Synchrotron Radiation Research Institute, Mikazuki~679-5198, Japan}
\author{H.~Fujimura}
\affiliation{Laboratory of Nuclear Science, Tohoku University,Sendai~982-0826, Japan}
\author{M.~Fujiwara}
\affiliation{Research Center for Nuclear Physics, Osaka University, Ibaraki~567-0047,
Japan}
\author{S.~Fukui}
\affiliation{Department of Physics and Astrophysics, Nagoya University, Nagoya, Aichi~464-8602, Japan}
\author{H.~Hasegawa}
\affiliation{Kansai Photon Science Institute, Japan Atomic Energy Agency, 619-0215
Kizu, Japan}
\author{K.~Hicks}
\affiliation{Department of Physics and Astronomy, Ohio University, Athens, 
Ohio~45701, USA} 
\author{K.~Horie} 
\affiliation{Department of Physics, Osaka University, Toyonaka~560-0043, Japan} 
\author{T.~Hotta} 
\affiliation{Research Center for Nuclear Physics, Osaka University, Ibaraki~567-0047, Japan} 
\author{K.~Imai}
\affiliation{Department of Physics, Kyoto University, Kyoto~606-8502, Japan}
\author{T.~Ishikawa}
\affiliation{Laboratory of Nuclear Science, Tohoku University,Sendai~982-0826, Japan}
\author{T.~Iwata}
\affiliation{Department of Physics, Yamagata University, Yamagata 990-8560, Japan}
\author{Y.~Kato}
\affiliation{Research Center for Nuclear Physics, Osaka University, Ibaraki~567-0047, Japan} 
\author{H.~Kawai}
\affiliation{Graduate School of Science and Technology, Chiba University,
Chiba~263-8522, Japan}
\author{Z.Y.~Kim}
\affiliation{School of Physics, Seoul National University, Seoul, 151-747, Korea}
\author{K.~Kino}
\affiliation{Research Center for Nuclear Physics, Osaka University, Ibaraki~567-0047, Japan} 
\author{H.~Kohri}
\affiliation{Research Center for Nuclear Physics, Osaka University, Ibaraki~567-0047, Japan} 
\author{N.~Kumagai}
\affiliation{Japan Synchrotron Radiation Research Institute,
Sayo, Hyogo~679-5198, Japan}
\author{S. Makino}
\affiliation{Wakayama Medical University, Wakayama, Wakayama 641-8509, Japan}
\author{T.~Matsuda}
\affiliation{Department of Applied Physics, Miyazaki University, Miyazaki 889-2192,
Japan}
\author{N.~Matsuoka}
\affiliation{Research Center for Nuclear Physics, Osaka University, Ibaraki~567-0047, 
Japan}
\author{T.~Matsumura}
\affiliation{Department of Applied Physics, National Defense Academy, Yokosuka~239-8686, Japan}
\author{T.~Mibe}
\affiliation{High Energy Accelerator Reseach Organization, KEK, 1-1 Oho Tsukuba, Ibaraki~305-0801, Japan}
\author{M.~Miyabe}
\affiliation{Department of Physics, Kyoto University, Kyoto~606-8502, Japan}
\author{Y.~Miyachi}
\affiliation{Department of Physics, Tokyo Institute of Technology, Tokyo 152-8551,
Japan}
\author{M.~Niiyama}
\affiliation{The Institute of Physical and Chemical Research, Wako, Saitama}
\author{M.~Nomachi}
\affiliation{Department of Physics, Osaka University, Toyonaka~560-0043, Japan}
\author{Y.~Ohashi}
\affiliation{Japan Synchrotron Radiation Research Institute, Mikazuki~679-5198, Japan} 
\author{H.~Ohkuma}
\affiliation{Japan Synchrotron Radiation Research Institute, Mikazuki~679-5198, Japan}
\author{T.~Ooba}
\affiliation{Graduate School of Science and Technology, Chiba University,
Chiba~263-8522, Japan}
\author{D.S.~Oshuev}
\affiliation{Institute of Physics, Academia Sinica, Taipei~11529, Taiwan}
\author{C.~Rangacharyulu}
\affiliation{Department of Physics and Engineering Physics, University of Saskatchewan,
Saskatoon, Saskatchewan~S7N5E2, Canada}
\author{A.~Sakaguchi}
\affiliation{Department of Physics, Osaka University, Toyonaka~560-0043, Japan}
\author{P.M.~Shagin}
\affiliation{School of Physics and Astronomy, University of Minnesota, Minneapolis, Minnesota 55455}
\author{Y.~Shiino}
\affiliation{Graduate School of Science and Technology, Chiba University,
Chiba~263-8522, Japan}
\author{A.~Shimizu}
\affiliation{Research Center for Nuclear Physics, Osaka University, Ibaraki~567-0047, 
Japan}
\author{H.~Shimizu}
\affiliation{Laboratory of Nuclear Science, Tohoku University,Sendai~982-0826, Japan}
\author{Y.~Sugaya}
\affiliation{Department of Physics, Osaka University, Toyonaka~560-0043, Japan} 
\author{M.~Sumihama}
\affiliation{Research Center for Nuclear Physics, Osaka University, Ibaraki~567-0047, Japan} 
\author{Y.~Toi}
\affiliation{Department of Applied Physics, Miyazaki University, Miyazaki 889-2192,
Japan}
\author{H.~Toyokawa}
\affiliation{Japan Synchrotron Radiation Research Institute,
Sayo, Hyogo~679-5198, Japan}
\author{A.~Wakai}
\affiliation{Akita Research Institute of Brain and Blood Vessels, Akita 010-0874, Japan}
\author{C.W.~Wang}
\affiliation{Institute of Physics, Academia Sinica, Taipei~11529, Taiwan}
\author{S.C.~Wang}
\affiliation{Institute of Physics, Academia Sinica, Taipei~11529, Taiwan}
\author{K.~Yonehara}
\affiliation{Illinois Institute of Technology, Chicago, Illinois 60616, USA}
\author{T.~Yorita}
\affiliation{Research Center for Nuclear Physics, Osaka University, Ibaraki~567-0047, Japan} 
\author{M.~Yoshimura}
\affiliation{Institute for Protein Research, Osaka Univirsity, Osaka 565-0871, Japan}
\author{M.~Yosoi}
\affiliation{Research Center for Nuclear Physics, Osaka University, Ibaraki~567-0047, Japan} 
\author{R.G.T.~Zegers}
\affiliation{National Superconducting Cyclotron Laboratory,
Michigan State University, Michigan 48824, USA}



\begin{abstract}
The $\gamma d \to K^+K^-pn$ reaction has been studied to search for
the evidence of the $\Theta^+$ by detecting $K^+K^-$ pairs at forward
angles.  The Fermi-motion corrected $nK^+$ invariant mass distribution
shows a narrow peak at $1.524 \pm 0.002 + 0.003$ GeV/$c^2$. The
statistical significance of the peak calculated from a shape analysis
is 5.1 $\sigma$, and the differential cross-section for the $\gamma n
\to K^- \Theta^+$ reaction is estimated to be $12 \pm 2$ nb/sr in the
photon energy range from 2.0 GeV to 2.4 GeV in the LEPS angular range
by assuming the isotropic production of the $\Theta^+$ in the $\gamma
n$ center-of-mass system.  The obtained results support the existence
of the $\Theta^+$.
\end{abstract}


\maketitle

\section{INTRODUCTION}

Since the LEPS collaboration reported the observation of a narrow baryon 
resonance-like structure in the $nK^{+}$ invariant mass spectrum produced in 
$\gamma n\to K^+K^-n$ reactions \cite{LEPS}, a considerable number of experiments have 
been carried out to check the existence of the exotic baryon, now called the $\Theta 
^{+}$. The $\Theta ^{+}$ is a genuine exotic baryon with the minimum 
quark configuration of $uudd\overline s $, for which a narrow decay width 
and a light mass were first predicted by Diakonov, Petrov, and Polyakov 
using a chiral quark soliton model~\cite{Diakonov}. 
Although the LEPS result seemed to be 
supported by several experiments which reported positive evidence for the 
existence of the $\Theta ^{+}$ in various reactions
\cite{DIANA, CLAS, SAPHIR, CLAS2, nu, HERMES, ZEUS, COSY, SVD}, 
the experimental situation soon became controversial. 

Many experiments at the high energy, especially collider experiments, found 
no positive evidence in the $pK_{s}$ invariant mass distributions with a 
good mass resolution and high statistics
\cite{HyperCP, HERA-B, ALEPH, BES, BABAR, CDF, SPHINX}. 
A typical upper limit for the 
inclusive production rate for the $\Theta ^{+}$ is less than 1{\%} of that 
of the $\Lambda $(1520). The production mechanism of the $\Theta ^{+ 
}$ might be very different from those of ordinary baryons if the $\Theta 
^{+}$ exists~\cite{Titov2004}.

The CLAS collaboration searched for the $\Theta ^{+}$ in the $\gamma
p\to \overline K ^0K^+n$ reaction in the photon energy range from 1.6
GeV to 3.8 GeV with an integrated luminosity of 70
pb$^{-1}$~\cite{DeVita2006}.  The upper limit for the $\gamma p\to
\overline K ^0\Theta ^+$ reaction was determined to be 0.7 nb. The
non-observation of the $\Theta^+$ might be explained by a weak
$K^{\ast }N\Theta ^{+}$ coupling~\cite{Azimov:2006}. On the other
hand, if the $K^{\ast }$ coupling constant is small, the
photo-production cross-section of the $\Theta ^{+}$ from a proton
could be much smaller than that from a neutron~\cite{Nam2006}.

The experiment which is most relevant to the current study was also
carried out by the CLAS collaboration~\cite{McKinnon2006}.  The search
was done by detecting all charged particles in the final state in
$\gamma d\to pK^-K^+n$ reactions with one order of magnitude higher
statistics than the previous experiment~\cite{CLAS}. The neutron
momentum was reconstructed by using the missing momentum technique,
and the $\Theta ^{+}$ was searched in the $nK^{+}$ invariant
mass distribution. No narrow peak was observed, and the upper limit
(95 {\%} CL) for the elementary $\gamma n\to K^-\Theta ^+$ reaction
was obtained to be $\sim$ 3nb by using a phenomenological model based
on the $\Lambda $(1520) production to estimate the probability that the
spectator proton is re-scattered and gains enough energy to be
detected by the CLAS detector.

Other dedicated experiments using $\pi $~\cite{Miwa2006},
$K$~\cite{Miwa2007}, and proton~\cite{AbdelBary2006} beams have also
shown no evidence for the $\Theta ^{+}$ production. Although the
theoretical interpretation of those null results contains some
uncertainties due to model dependences in the cross-section
calculations, a strong $K^{\ast }N\Theta ^{+}$ coupling is
unlikely if the $\Theta ^{+}$ exists.

A model independent determination of the $\Theta ^{+}$ width is possible
by investigating the reverse reaction of the $\Theta ^{+}$ decay:
$K^+n\to \Theta ^+$. The DIANA collaboration observed evidence of
the $\Theta ^{+}$ in the $pK_S$ invariant mass distribution from
$K^+Xe\to K^0pX$ reactions in a bubble chamber. The $\Theta ^{+}$ width
was estimated to be $0.36 \pm 0.11$ MeV/c$^{2}$ from the production
cross-section~\cite{Barmin2007}.  This result is barely consistent with the
90{\%}-CL upper limit of 0.64 MeV/$c^{2}$ from the Belle collaboration
obtained by analyzing events from secondary kaon interactions in the
material of the detector~\cite{Abe2005}.

To summarize the situation, if the $\Theta ^{+}$ exists, 1) its
production is highly reaction dependent, 2) the coupling to $K~{\ast}N$ must
be small, and 3) the decay width must be less than 1
MeV/$c^{2}$. Thus, it is desirable to study reactions which are
sensitive to the $KN\Theta ^{+}$ coupling. The quasi-free
reaction $\gamma n\to K^-\Theta ^+$ is one of such reactions since a
$\gamma $ can couple to a $K^+K^-$pair.

In this paper we present a study of the photo-production of the
$\Theta ^{+}$ from a neutron by closely comparing it with the
photo-production of the $\Lambda(1520)$ from a proton in a
deuteron. Because the LEPS detector has a symmetric acceptance
for positive and negative particles, a similar procedure can
be applied to the both analyses. The validity of corrections and
event selection criteria can also be cross-checked.   

The analysis is performed using the data collected with the
LEPS detector in 2002-2003, where the statistics has been improved
by a factor of 8 over the previous measurement~\cite{LEPS}. 

\section{LEPS BEAMLINE AND DETECTOR}
A photon beam in the energy range from 1.5 GeV to 2.4 GeV is produced at 
SPring-8 by Compton back-scattering of laser photons from 8 GeV electrons in 
the storage ring. The energy of a scattered photon is measured by tagging 
the electron which is associated with the Compton scattering event by event. 
The energy resolution for the tagged photon is 10 MeV, and typical beam 
intensity with a 351-nm Ar laser is 10$^{6}$ photons/sec. The photons are 
alternatively injected into liquid deuterium (LD$_{2}$) or liquid hydrogen 
(LH$_{2}$) targets in a 16-cm long cell made of aluminum. The height of the 
interior of the cell is 60 mm, and the width is 40 mm at the entrance and 
100 mm at the exit. The windows of the cell are made of a Kapton polyimide 
film of 125 $\mu$m in thickness. 

The LEPS detector is a forward magnetic spectrometer which consists of a 
start counter (SC), a silicon vertex counter, an aerogel Cherenkov counter 
(AC), three drift chambers, a dipole magnet, and a wall of time-of-flight 
scintillation counters (TOF). The aperture of the 0.7-T dipole magnet is 55 
cm high and 135 cm wide. The pole length is 60 cm. The angular coverage of 
the spectrometer is approximately $\pm 20$ and $\pm 10$ degrees in the 
horizontal and the vertical directions, respectively. The distance from the 
SC to the TOF is 4 m. A typical momentum resolution, $\Delta p/p$, for 
a charged particle with $p=1.0$ GeV/$c$ is 0.6 {\%}, and the TOF resolution is 
140 ps. The details of the detector and the quality of the
particle identification are described elsewhere~\cite{Sumihama2006}. 

The event trigger requires a coincidence of signals from the SC and TOF. A 
particle with $p<0.3$ GeV/$c$ cannot reach the TOF. Signals from the AC are 
used to veto events with pair-created $e^+e^-$ or a pion with $p>0.6$ GeV/$c$ online. 
A typical trigger rate is 90 Hz for the LD$_{2}$ runs and 60 Hz for the 
LH$_{2}$ runs. 

The integrated numbers of photons in the energy range $1.5-2.4$ GeV and 
$2.0-2.4$ GeV were $3.93\times 10^{12}$ and $2.15\times 10^{12}$ for the 
LD$_{2 }$ target runs, and the corresponding numbers were $2.52\times 
10^{12}$ and $1.34\times 10^{12}$ for the LH$_{2}$ target runs. The total 
numbers of events collected with the LD$_{2 }$ target and the LH$_{2 }$ target 
were $4.5\times 10^8$ and $2.2\times 10^8$, respectively. Events in the
energy range $2.0-2.4$ GeV have been used for the current study.

\section{EVENT SELECTION}
We select events of the type $\gamma d\to K^+K^-X$, where $X$ 
denotes particles which are not required to be identified by the LEPS 
detector. 

The momentum of a charged particle is reconstructed from the track 
information, and the velocity is obtained from the track length and the TOF 
information. The mass of the charged particle is calculated from the 
reconstructed momentum and velocity. The momentum dependent mass resolution 
$\sigma _{M}$ for a kaon is calculated by using the measured momentum and 
TOF resolutions. A charged particle is identified as a kaon if the 
reconstructed mass is within 3.5$\sigma _{M}$ of the nominal value. Events 
with a $K^+K^-$ pair are selected, and the vertex point of the two kaon 
tracks is required to be within the target volume. A hit position of a track 
in the vertical direction at the TOF wall is reconstructed from the charge 
ratio and time difference of the signals from both ends of a TOF 
counter, and the horizontal position is obtained from the counter 
segmentation. The consistency between the reconstructed hit position and the 
extrapolated track at the TOF wall is checked to remove events with a 
decay-in-flight kaon. A total of 25820 and 8675 events passed all the 
selection cuts for the LD$_{2}$ runs and LH$_{2}$ runs, respectively. 

\begin{figure}[htbp]
\begin{tabular}{c c}
\begin{minipage}{0.5\hsize}
\begin{center}
 \includegraphics[width=7.5cm,height=7.5cm,keepaspectratio]
{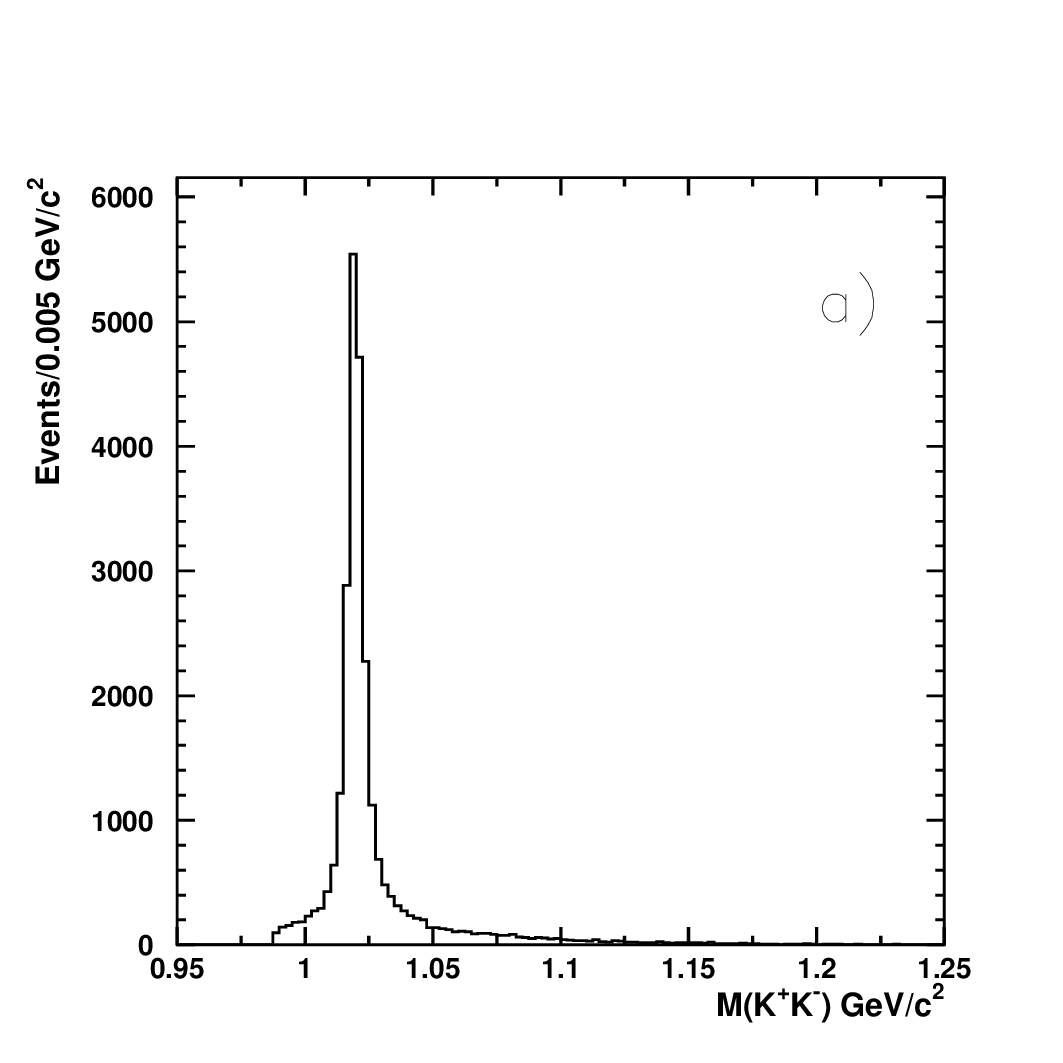}
\end{center}
\end{minipage} & %
\begin{minipage}{0.5\hsize}
\begin{center}
 \includegraphics[width=7.5cm,height=7.5cm,keepaspectratio]
{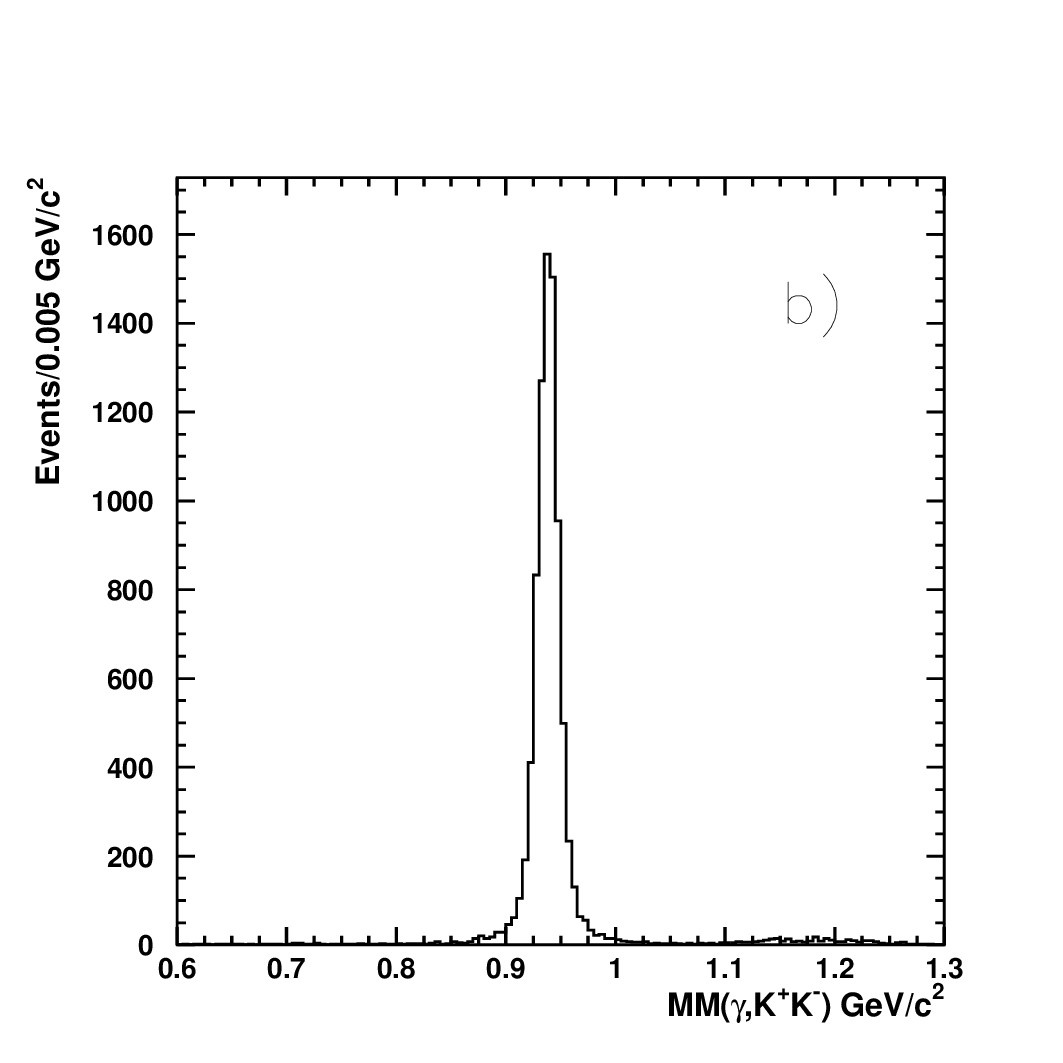}
\end{center}
\end{minipage} \\
\begin{minipage}{0.5\hsize}
\begin{center}
 \includegraphics[width=7.5cm,height=7.5cm,keepaspectratio]
{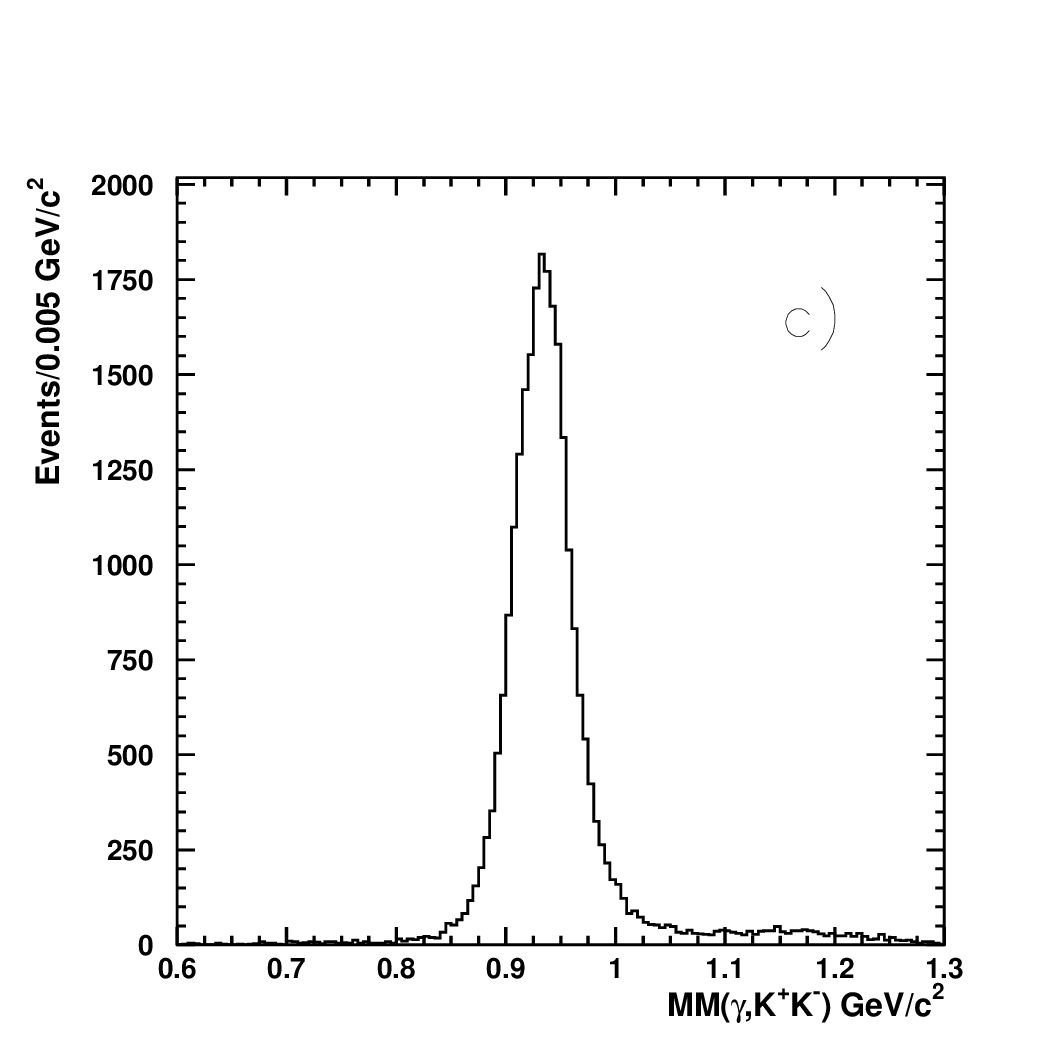}
\end{center}
\end{minipage} & %
\begin{minipage}{0.5\hsize}
\begin{center}
 \includegraphics[width=7.5cm,height=7.5cm,keepaspectratio]
{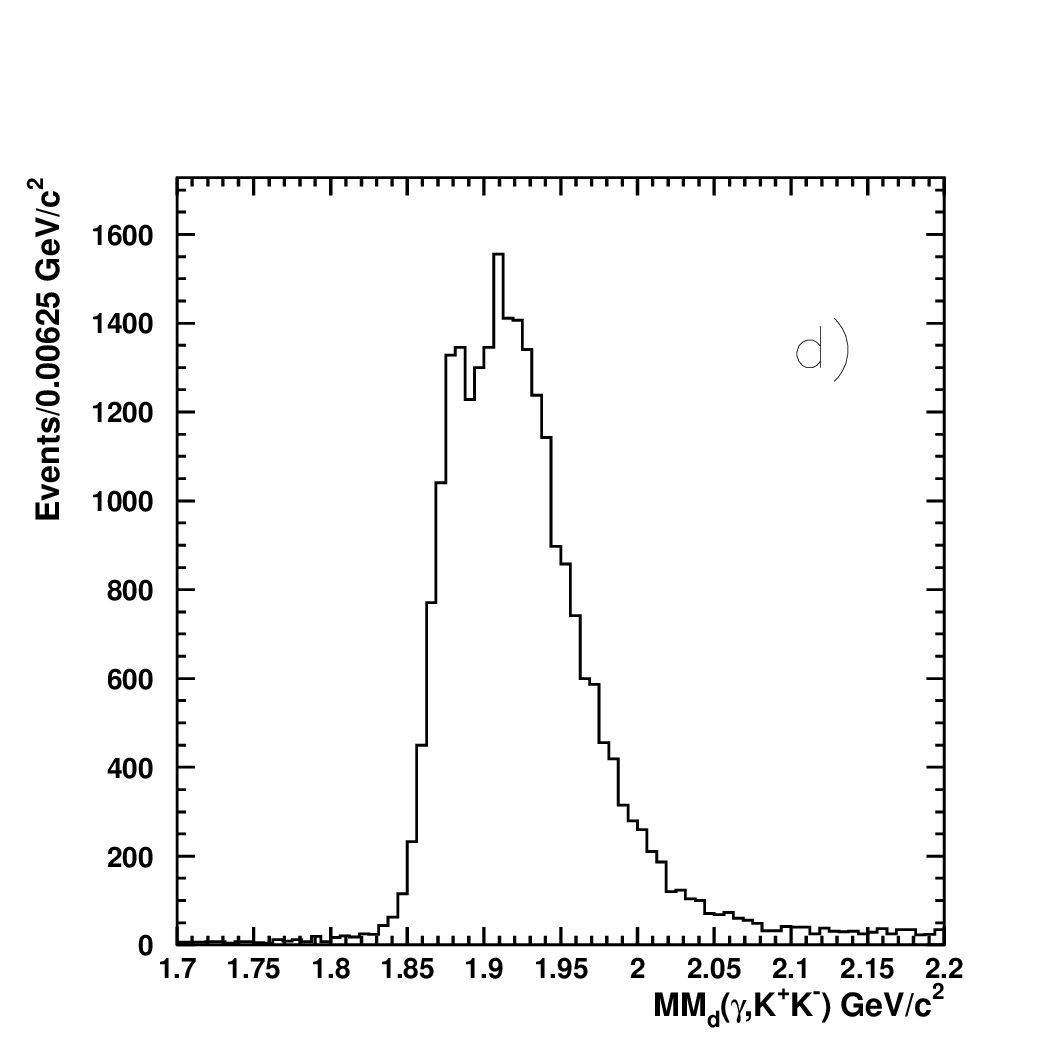}
\end{center}
\end{minipage} \\
\end{tabular}
\caption{
(a) $M(K^+K^-)$ distribution for the LD$_{2}$ runs.
(b) $MM(\gamma ,K^+K^-)$ distribution for the LH$_{2}$ runs.
(c) $MM(\gamma,K^+K^-)$ distribution for the LD$_2$ runs.
(d) $MM_d (\gamma,K^+K^-))$ distribution for the LD$_2$ runs.
}
\label{fig:ev}
\end{figure}

The invariant $K^+K^-$ mass ($M(K^+K^-)$) distribution for the
LD$_{2}$ runs is shown in Fig.~\ref{fig:ev}(a).  A narrow peak at 1.02
GeV/$c^{2}$ is due to $\phi \to K^+K^-$ decays. Events with 1.01
GeV/$c^{2}$ $< M(K^+K^-) <$ 1.03 GeV/$c^{2}$ account for approximately
74{\%} of the $K^+K^-$ events. The $p(\gamma ,K^+K^-)$ missing mass
($MM(\gamma ,K^+K^-))$ distribution for the LH$_{2}$ runs is shown in
Fig.~\ref{fig:ev}(b).  These events are dominated by elastic processes. The
missing mass resolution for a proton is seen to be 10 MeV/$c^{2}$. Inelastic
events with a high missing mass value of $MM(\gamma ,K^+K^-)>1.08$
GeV/$c^{2}$ are 3.5 {\%} of the selected $K^+K^-$ events.  Events due
to misidentification of a pion pair, which result in a low missing
mass value, are estimated to be less than 1 {\%}. The $MM(\gamma
,K^+K^-)$ distribution for the LD$_2$ runs is shown in
Fig.~\ref{fig:ev}(c). The struck nucleon in the initial state has been
assumed to be at rest. The peak near the nucleon mass is wide due to
the Fermi motion of the nucleon. Some of the events in the lower tail
region are due to coherent processes~\cite{Chang2008}, which are
identified as a small peak at 1.88 GeV/$c^2$ in the $d(\gamma
,K^+K^-)$ missing mass ($MM_d (\gamma ,K^+K^-))$ distribution as shown
in Fig.~\ref{fig:ev}(d).

\section{MINIMUM MOMENTUM SPECTATOR APPROXIMATION}
Because the momenta of the target nucleons are not measured, some approximation is 
necessary to obtain the invariant mass of $pK^-$ or $nK^+$ pairs from $\gamma 
d\to K^+K^-pn$ reactions. The processes of interest are sequential processes 
of quasi-free productions of $\Lambda (1520)$ or $\Theta ^+$ and their 
decays; $\gamma p\to K^+\Lambda (1520)\to K^+K^-p$ and $\gamma n\to 
K^-\Theta ^+\to K^-K^+n$. We call the remaining nucleon which is not associated 
with the quasi-free processes a spectator. The spectator momentum due to the 
Fermi motion is approximately $\sim $80 MeV/$c$ for a deuteron. And it is small 
compared with momenta of a photon and kaons which are detected by the LEPS 
spectrometer. Therefore, the simplest approximation is to neglect the 
existence of a spectator. In this case the $nK^+$ invariant mass is obtained 
by calculating a $(\gamma ,K^-)$ missing mass ($MM(\gamma ,K^-))$ with the 
assumption that the struck neutron is at rest in the initial state and 
always on-shell. We call this approximation the free nucleon approximation 
(FNA). A Monte-Carlo simulation study shows the mass resolution of the 
$\Theta ^{+}$ using the FNA is $\sim $30 MeV/$c^{2}$, which is mainly
determined by the Fermi motion of a neutron. 

The minimum momentum spectator approximation (MMSA) has been developed
in order to improve the mass resolution. In the MMSA a spectator is
assumed to have the minimum momentum for the given total
energy-momentum ($p_{pn} =(E_{pn},\overrightarrow p _{pn} )$) of 
a $pn$ pair, which is in turn
assumed to be equal to the missing energy-momentum of the $\gamma d\to
K^+K^-X$ reaction:
\begin{eqnarray}
p_{pn} =p_{miss} =p_{_\gamma } +p_d -p_{K^+} -p_{K^-}.
\end{eqnarray}
This assumption is not valid for inelastic events
with an additional pion. However, these events can be removed easily
as it will be shown below. 
Note that $p_{pn}$ is derived from measured quantities and 
the deuteron mass ($p_d =(m_d ,0))$. The magnitude of the nucleon momentum ($p_{CM} 
)$ in the $pn$ center-of-mass system is then given by
\begin{eqnarray}
p_{CM} =\frac{\sqrt {(M_{pn} +m_p +m_n )(M_{pn} -m_p +m_n )(M_{pn} +m_p -m_n 
)(M_{pn} -m_p -m_n )} }{2M_{pn} }
\end{eqnarray}
in terms of the proton mass ($m_p )$, the neutron mass ($m_n$ ), and the 
invariant mass of a $pn$ pair ($M_{pn}^2 =p_{pn}^2$ ). If, for a 
particular event, $M_{pn}$ is found to be less than $m_p + m_n$ due to finite detector resolutions and 
coherent contributions, it is set equal to $m_p + m_n$. 

The momentum of a nucleon in the laboratory system has the minimum magnitude 
if the direction is anti-parallel to that of the total missing momentum. This 
topology is assumed in the MMSA. The minimum momentum, $p_{\min } $, is 
defined as the component of the spectator momentum in the direction of the 
missing momentum. Thus, it is given by
\begin{eqnarray}
p_{\min } =-p_{CM} \cdot \frac{E_{miss} }{M_{pn} }+\sqrt {p_{CM}^2 +m_N^2 } 
\cdot \frac{\left| {\overrightarrow p _{miss} } \right|}{M_{pn} },
\end{eqnarray}
where $m_N$ is the mass of a nucleon which is assumed to be a spectator. 
With this approximation, the momentum component of the other nucleon in the 
direction of the missing momentum is given by
\begin{eqnarray}
p_{res} =\left| {\overrightarrow p _{miss} } \right|-p_{\min } .
\end{eqnarray}
If we assume that a spectator is a proton, the momentum of a neutron in the 
final state is given by
\begin{eqnarray}
\overrightarrow p _n =p_{res} \cdot \frac{\overrightarrow p _{miss} }{\left| 
{\overrightarrow p _{miss} } \right|}.
\end{eqnarray}
The invariant mass of the $nK^+(M(nK^+))$ is calculated by using the
above $\overrightarrow p _n $ and a measured $K^+$ momentum. The
resolution for the $\Theta ^+$ mass using the MMSA is 11 MeV/$c^{2}$,
which is an improvement over the FNA by a factor of 3.

Events which are not associated with quasi-free processes can be
identified from a large $\left| {p_{\min } } \right|_{ }$
value. Coherent processes which have a deuteron in the final state are
characterized by a positive $p_{\min } $ value which is approximately
equal to a half of $\left| {\overrightarrow p _{miss} } \right|$.
Inelastic reactions which create a pion in addition to a kaon pair
cause $p_{\min } $ to have a large negative value. Re-scattering
processes cause the $p_{\min } $ distribution to be dispersed. By
requiring $\left| {p_{\min } } \right|$ to be small, these background
events can be reduced.

\begin{figure}[htbp]
\begin{tabular}{c c}
\begin{minipage}{0.5\hsize}
\begin{center}
 \includegraphics[width=7.5cm,height=7.5cm,keepaspectratio]
{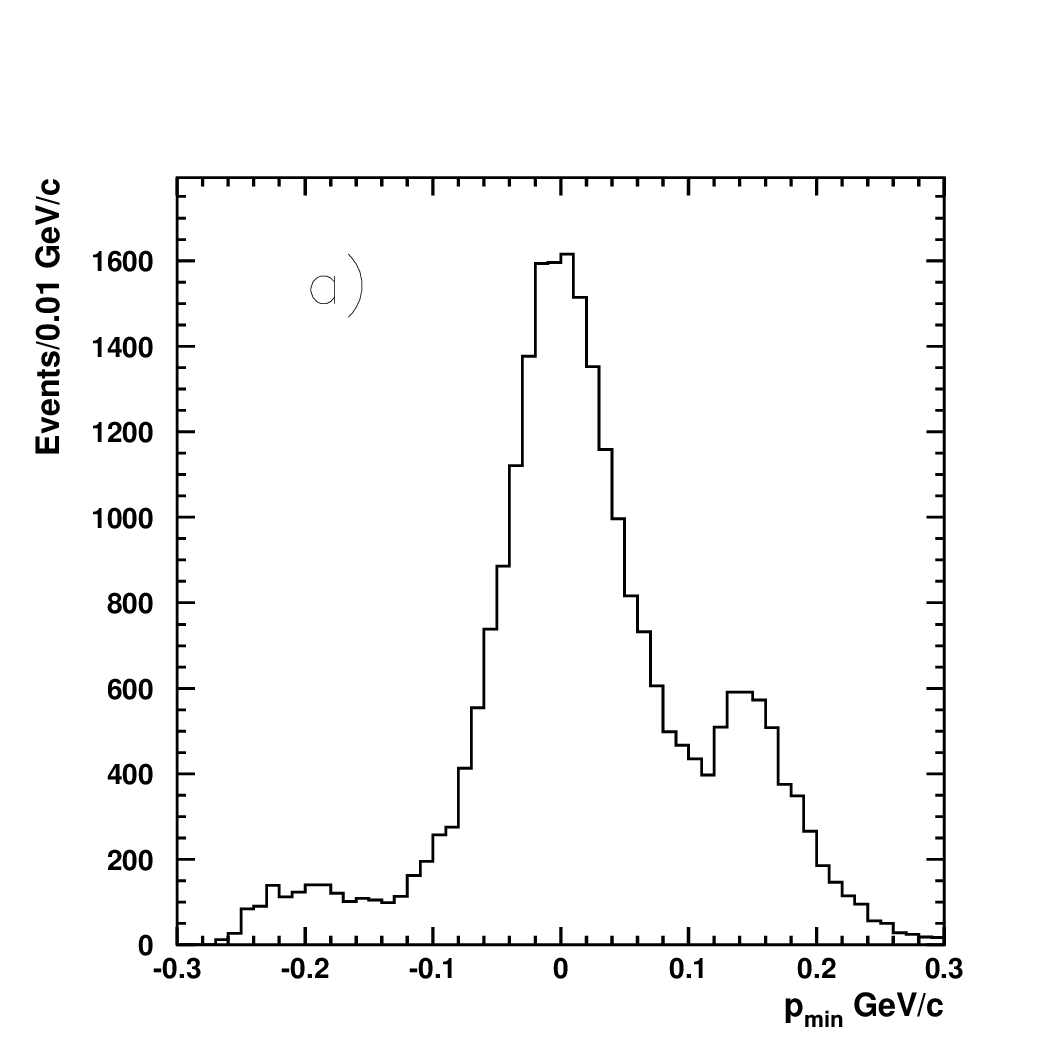}
\end{center}
\end{minipage} & %
\begin{minipage}{0.5\hsize}
\begin{center}
 \includegraphics[width=7.5cm,height=7.5cm,keepaspectratio]
{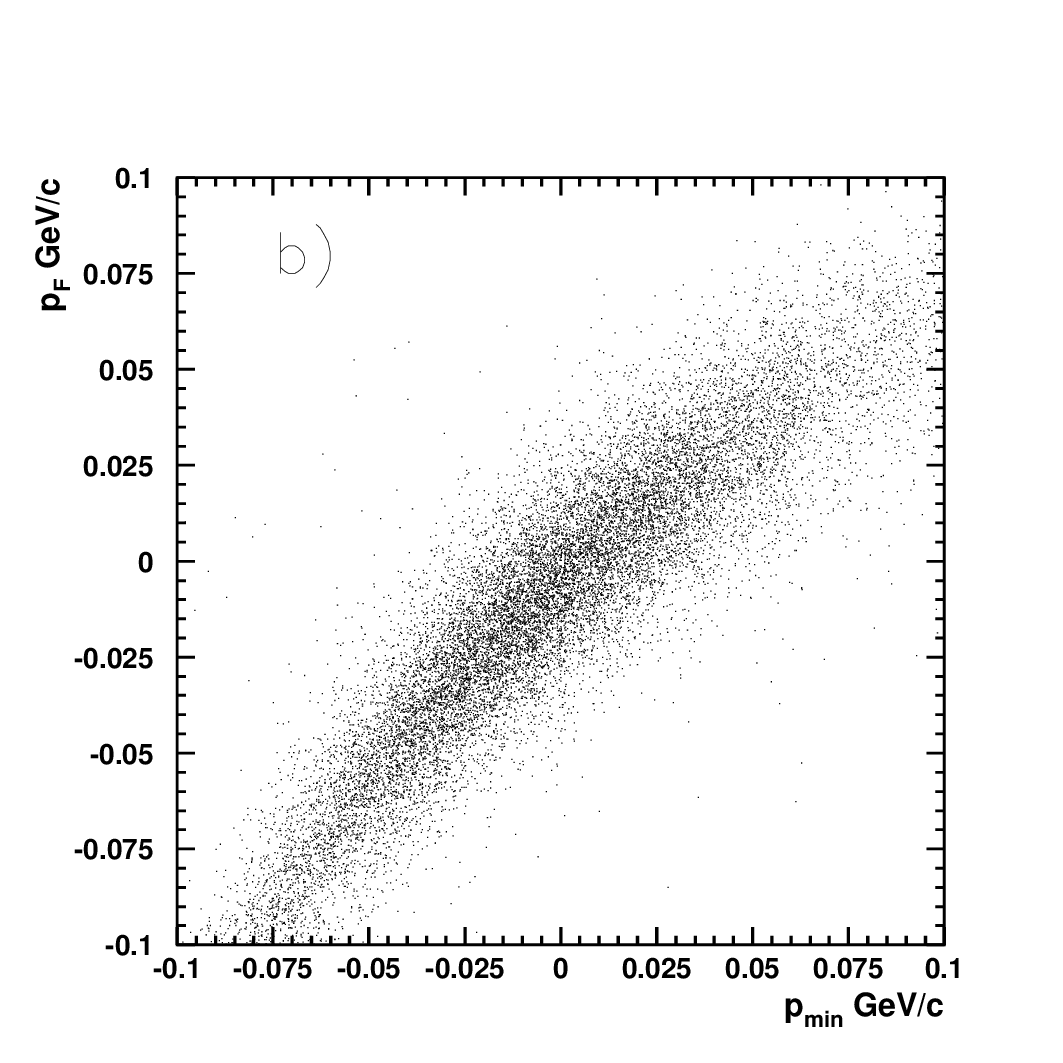}
\end{center}
\end{minipage} 
\end{tabular}
\caption{
(a) $p_{\min } $ distribution for the selected $K^+K^-$events.
(b) 2-d plot of $p_F$ \textit{vs.} $p_{\min}$ for non-resonant $K^+K^-$ MC events. }
\label{fig:pmin}
\end{figure}

The $p_{\min } $ distribution for the selected $K^+K^-$ events is
shown in Fig.~\ref{fig:pmin}(a). The main contribution from quasi-free
processes results in a peak near zero. The contribution from coherent
processes is seen as a bump near 0.15 GeV/$c$, and the inelastic
events concentrate in the region below -0.1 GeV/$c$. The projection of
the spectator momentum onto the axis of ${\overrightarrow p _{miss} }$
($p_F$) is well approximated by $p_{\min }$ as shown in
Fig.~\ref{fig:pmin}(b) for a Monte-Carlo simulation of non-resonant
$K^+K^-$ events.

\begin{figure}[htbp]
\begin{tabular}{c c}
\begin{minipage}{0.5\hsize}
\begin{center}
 \includegraphics[width=7.5cm,height=7.5cm,keepaspectratio]
{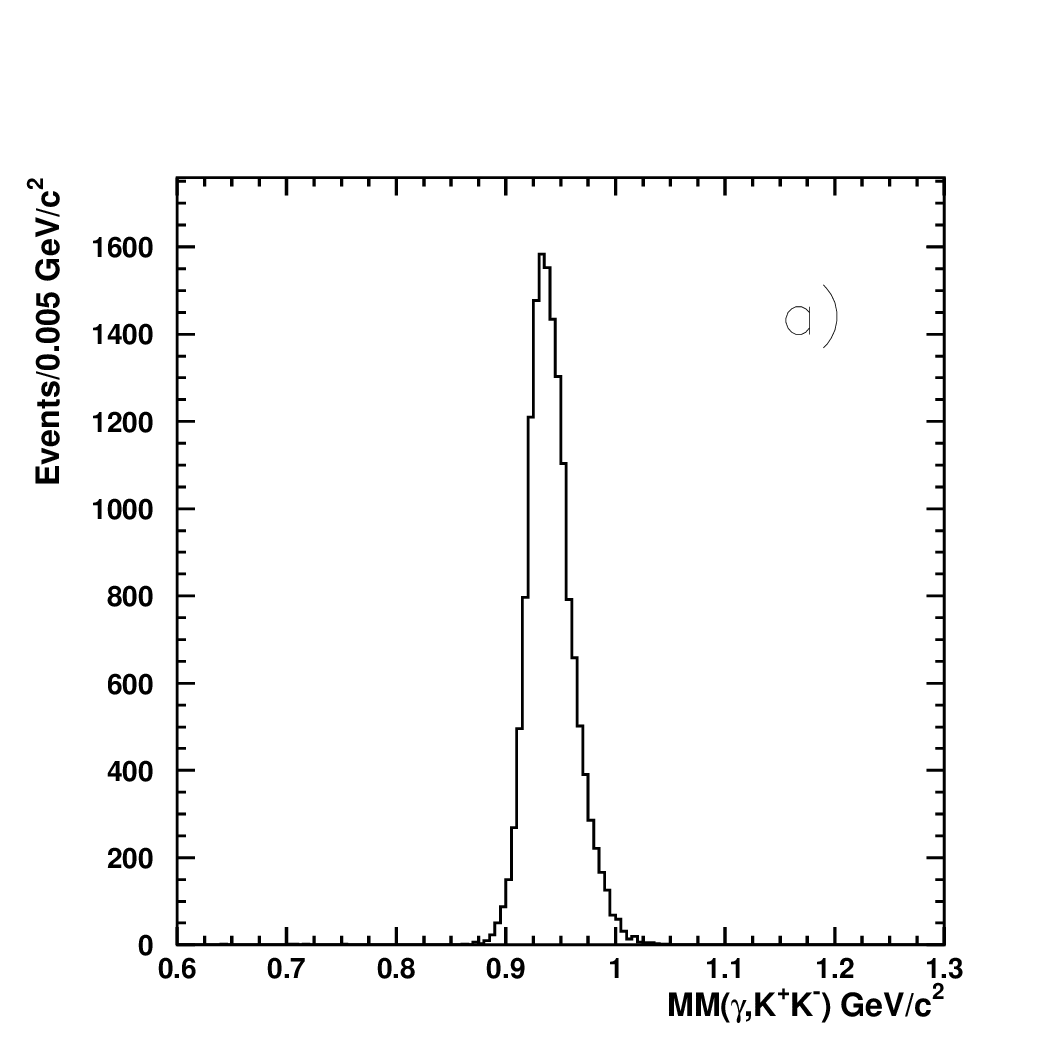}
\end{center}
\end{minipage} & %
\begin{minipage}{0.5\hsize}
\begin{center}
 \includegraphics[width=7.5cm,height=7.5cm,keepaspectratio]
{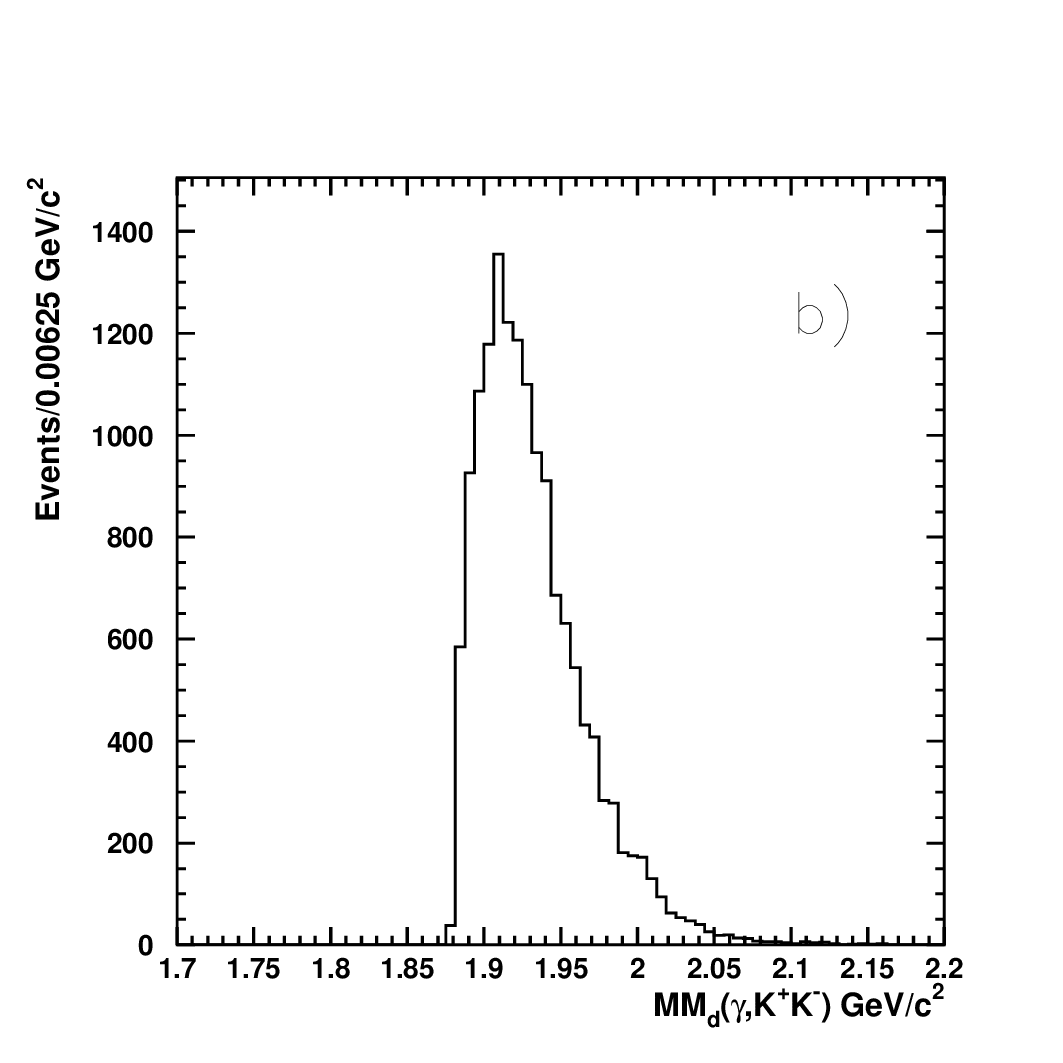}
\end{center}
\end{minipage}
\end{tabular}
\caption{
$MM(\gamma ,K^+K^-)$ (left) and $MM_d (\gamma ,K^+K^-)$ (right) distributions for
events with $\left| {p_{\min } } \right|<0.1$ GeV/$c$.
}
\label{fig:mm}
\end{figure}

The $MM(\gamma ,K^+K^-)$ and $MM_d (\gamma ,K^+K^-)$ distributions for 
events with $\left| {p_{\min } } \right|<0.1$ GeV/$c$ are shown in Fig.~\ref{fig:mm}. 
The inelastic and coherent contributions are successfully removed with 
$\left| {p_{\min } } \right|<0.1$ GeV/$c$.

Let $s$ be the square of the total center-of-mass energy of the $nK^+K^-$ 
system obtained with the MMSA. The effective photon energy $E_\gamma ^{eff} 
$is then defined by
\begin{eqnarray}
E_\gamma ^{eff} =\frac{s-m_n ^2}{2m_n }.
\end{eqnarray}
Note there is a one-to-one relation between $E_\gamma ^{eff}$ and $s$.
The $E_\gamma ^{eff}$ becomes close to $E_\gamma$ when the
magnitude of the Fermi momentum is small.  For the events with a small
$E_\gamma ^{eff} $, all of $M(K^+K^-)$, $M(nK^+)$, and $M(nK^-)$ have
a small value close to a threshold. Since we do not identify the
nucleon in the final state, $\Lambda (1520)$ events and $\phi $ events
from protons and neutrons may contribute in the small $M(nK^+)$ region
in this case.  Events with a large $E_\gamma^{eff}$ value are also
problematic because they are dominated by coherent events and events
with particle misidentifications.  Therefore, in addition to the
condition $\left| {p_{\min } } \right|<0.1$ GeV/$c$, events are
required to satisfy the condition 2.0 GeV $<E_\gamma ^{eff} <$ 2.5 GeV
for further analysis. The $\sqrt{s}$ value for $E_\gamma^{eff} =$ 2.0
GeV is 2.15 GeV. Thus, the maxmimum $M_{NK}$ is 1.65 GeV/$c^2$ at the
cut boundary.  Events with $E_\gamma^{eff} >$ 2.5 GeV have large Fermi
momentum for which the MMSA is not a good approximation.  The number
of events with $E_\gamma ^{eff} >$ 2.5 GeV is small (658 events)
compared to 14928 events with 2.0 GeV $<E_\gamma^{eff} <$ 2.5 GeV. For
the LH$_2$ runs, we require events to satisfy 2.0 GeV $<E_\gamma <$2.4
GeV and 0.9 GeV/$c^{2}<MM(\gamma ,K^+K^-)<$ 0.98 GeV/$c^{2}$. A total
of 6306 events have passed the requirements.

In principle, $m_n $ should be replaced by $m_p $ in the case of the 
$pK^+K^-$system. However, the difference between $E_\gamma ^{eff}$ values 
calculated with $m_n$ and $m_p$ is less than 1 MeV. Therefore, we use the 
mean of $m_p$ and $m_n$ for the calculation of $E_\gamma ^{eff} $.

\begin{figure}[htbp]
\begin{tabular}{c c}
\begin{minipage}{0.5\hsize}
\begin{center}
 \includegraphics[width=7.5cm,height=7.5cm,keepaspectratio]
{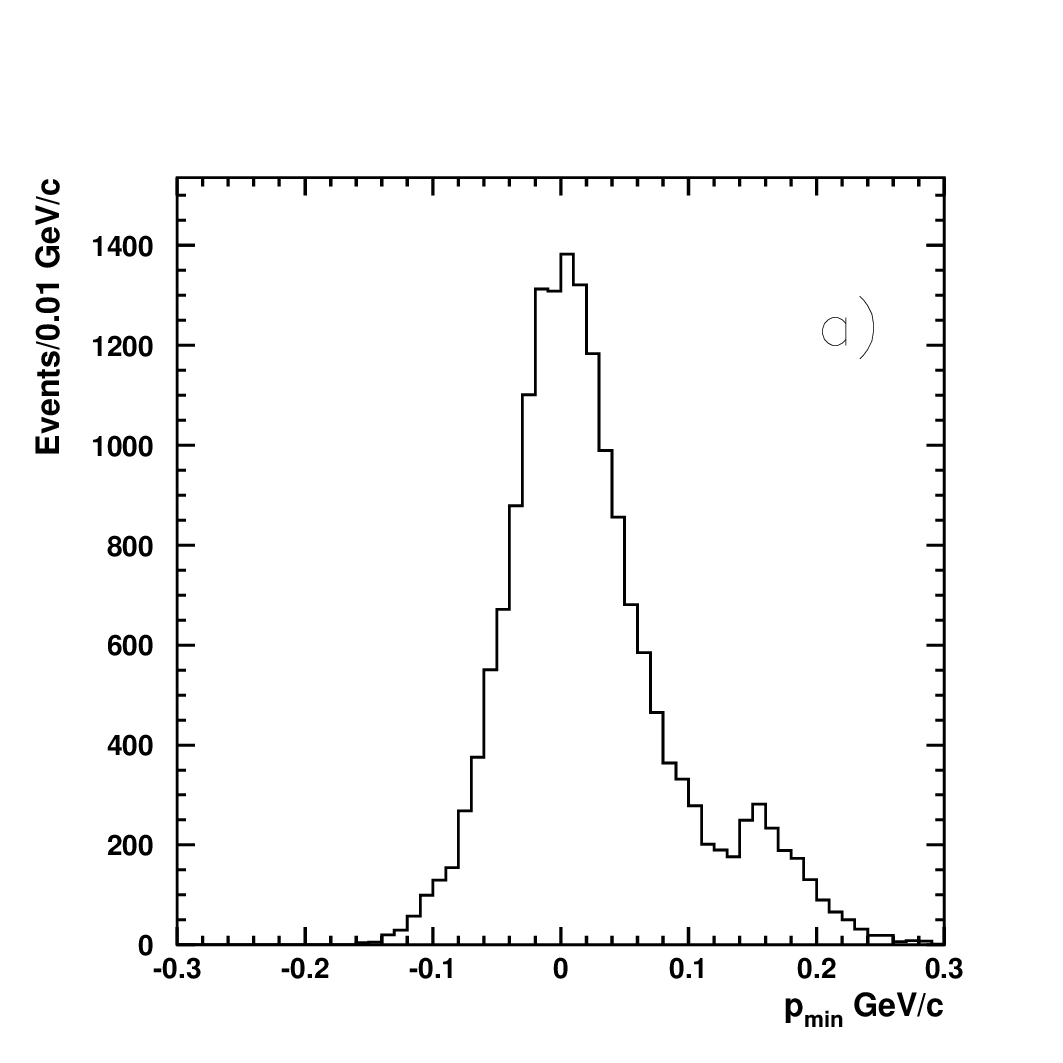}
\end{center}
\end{minipage} & %
\begin{minipage}{0.5\hsize}
\begin{center}
 \includegraphics[width=7.5cm,height=7.5cm,keepaspectratio]
{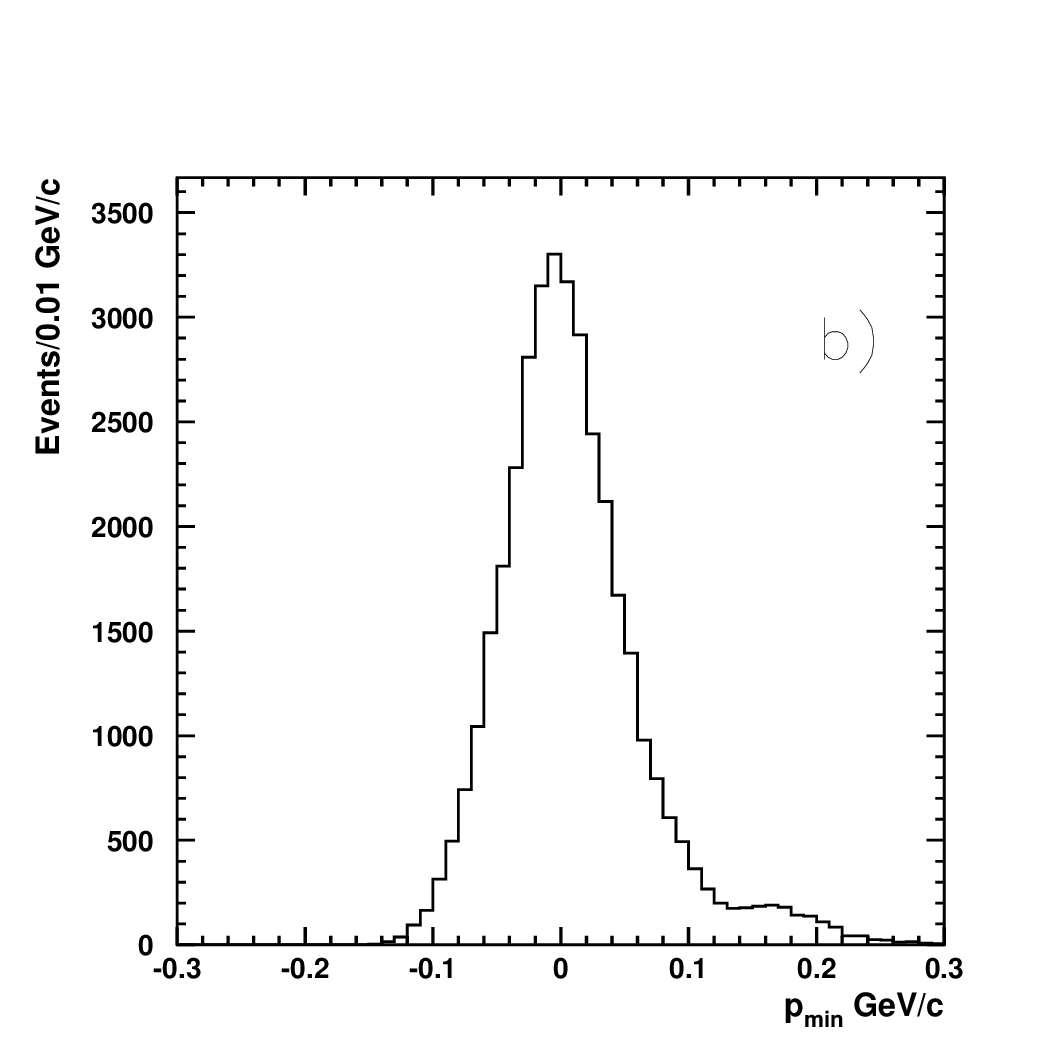}
\end{center}
\end{minipage}
\end{tabular}
\caption{
$p_{\min } $ distributions for events with 2.0 GeV $<E_\gamma ^{eff} <$
2.5 GeV for real data (left) and Monte-Carlo data (right).
}
\label{fig:pmin2}
\end{figure}

The $p_{\min } $ distributions for events with 2.0 GeV $<E_\gamma
^{eff} <$ 2.5 GeV are shown in Fig.~\ref{fig:pmin2}. Both the coherent
and inelastic events are strongly suppressed, and the main peak due to
quasi-free processes is well reproduced by a Monte-Carlo simulation
for non-resonant $\gamma n\to K^+K^-n$ reactions using a realistic
deuteron wave function~\cite{Lacombe}.  The non-uniform structure in
the higher tail of the distribution for the Monte-Carlo events is caused
by the special treatment of setting $M_{pn} = m_p + m_n$ when $M_{pn}$
becomes smaller than $m_p + m_n$ due to the finite resolutions.

In the Monte-Carlo study, the mass of a struck nucleon has been set to
be off-shell so that the total energy of two nucleons in the 
center-of-mass system is equal to $m_d$. The mass of a spectator
nucleon is always set to be on-shell.

\section{RANDOMIZED MINIMUM MOMENTUM METHOD}
\begin{figure}[htbp]
\begin{tabular}{c c}
\begin{minipage}{0.5\hsize}
\begin{center}
 \includegraphics[width=7.5cm,height=7.5cm,keepaspectratio]
{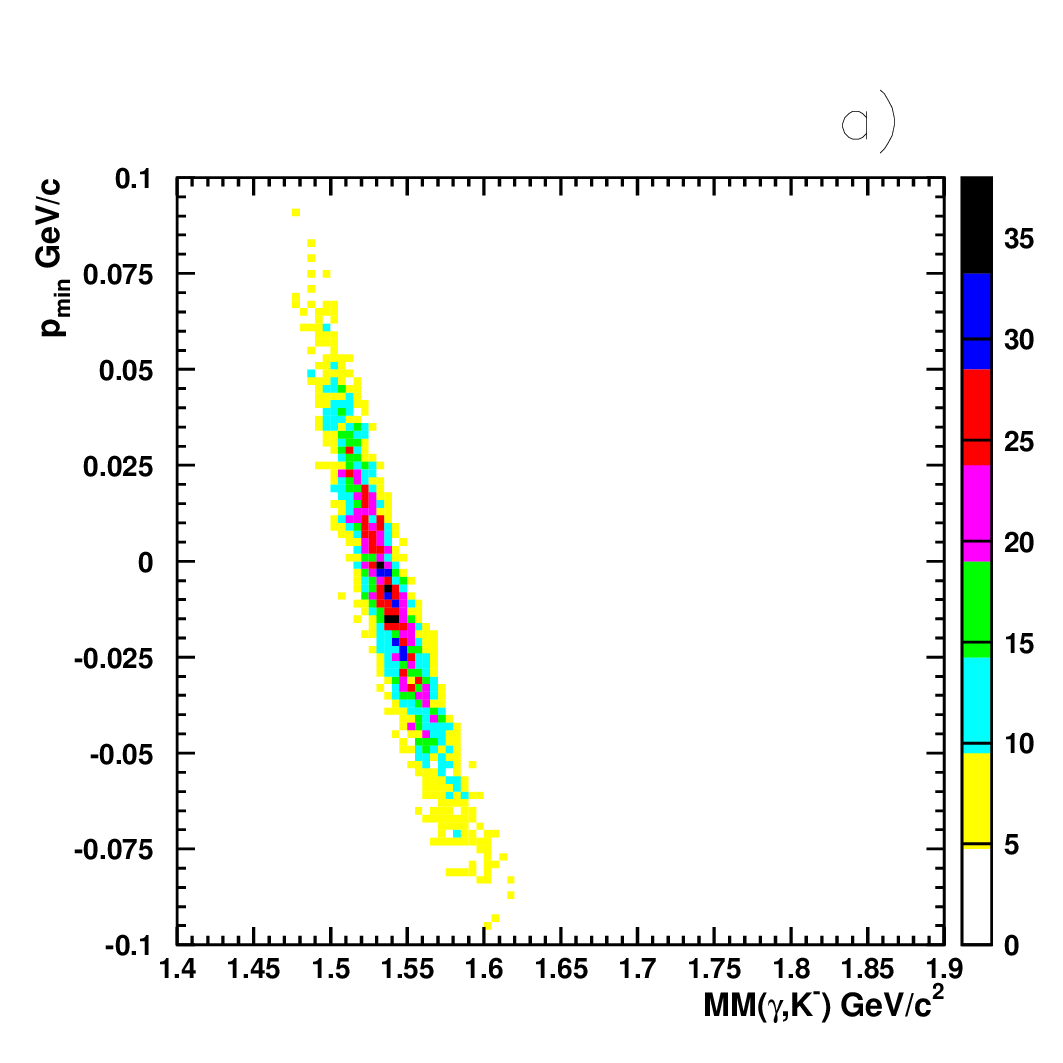}
\end{center}
\end{minipage} & %
\begin{minipage}{0.5\hsize}
\begin{center}
 \includegraphics[width=7.5cm,height=7.5cm,keepaspectratio]
{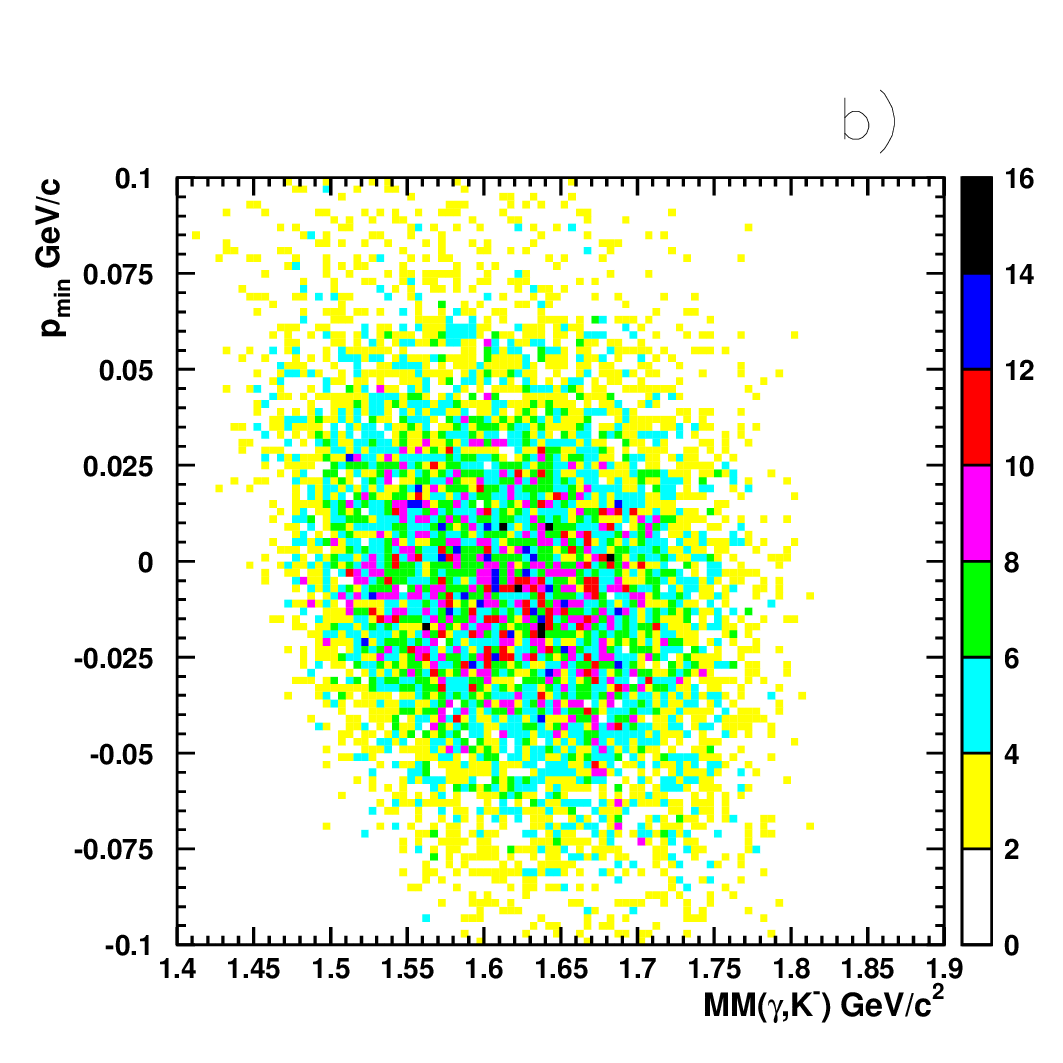}
\end{center}
\end{minipage}
\end{tabular}
\caption{
2-d plots of $p_{\min }$ \textit{vs.} $MM(\gamma ,K^-)$ for simulated signal 
($\Theta^+$) events (left) and non-resonant $K^+K^-$ events (right). 
}
\label{fig:pminvsmm}
\end{figure}

In this section, we develop a method to estimate the reasonable
$M(nK^+)$ spectrum shape for background contributions by using only
measured $E_{\gamma}$ and $\overrightarrow p _{K^-}$ values.  There is
a strong correlation between $p_{\min } $ and $MM(\gamma ,K^-)$ for
the signal Monte-Carlo events, while the correlation is very weak for
non-resonant background events as shown in Fig.~\ref{fig:pminvsmm}.
The nature of background events is characterized by the absence of
this correlation. Because $p_{\min } $ of a background event has a
random nature due to the Fermi motion, a reconstructed $p_{\min } $
can be replaced by a computer-generated one without changing the shape
of a $M(nK^+)$ distribution for background events.

\begin{figure}[htbp]
\begin{tabular}{c c}
\begin{minipage}{0.5\hsize}
\begin{center}
 \includegraphics[width=7.5cm,height=7.5cm,keepaspectratio]
{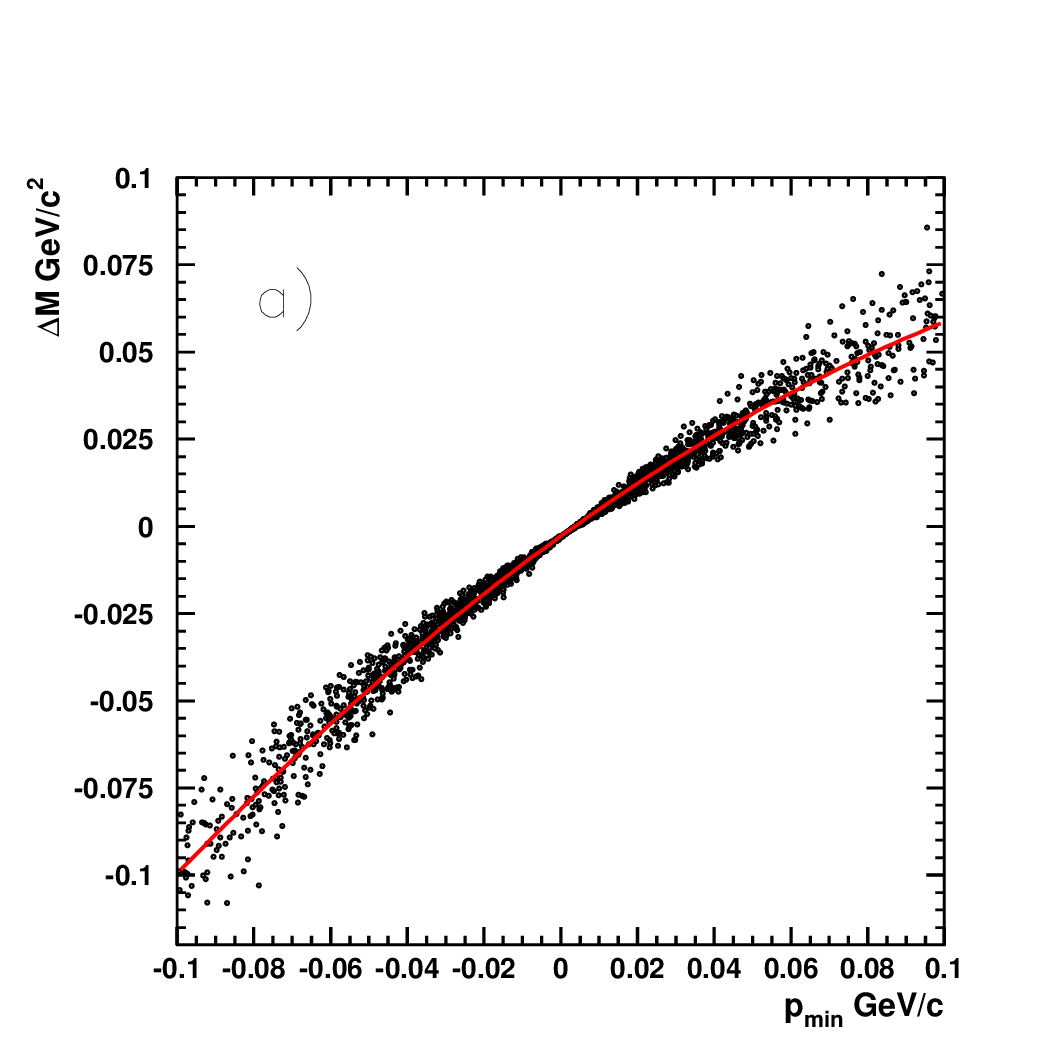}
\end{center}
\end{minipage} & %
\begin{minipage}{0.5\hsize}
\begin{center}
 \includegraphics[width=7.5cm,height=7.5cm,keepaspectratio]
{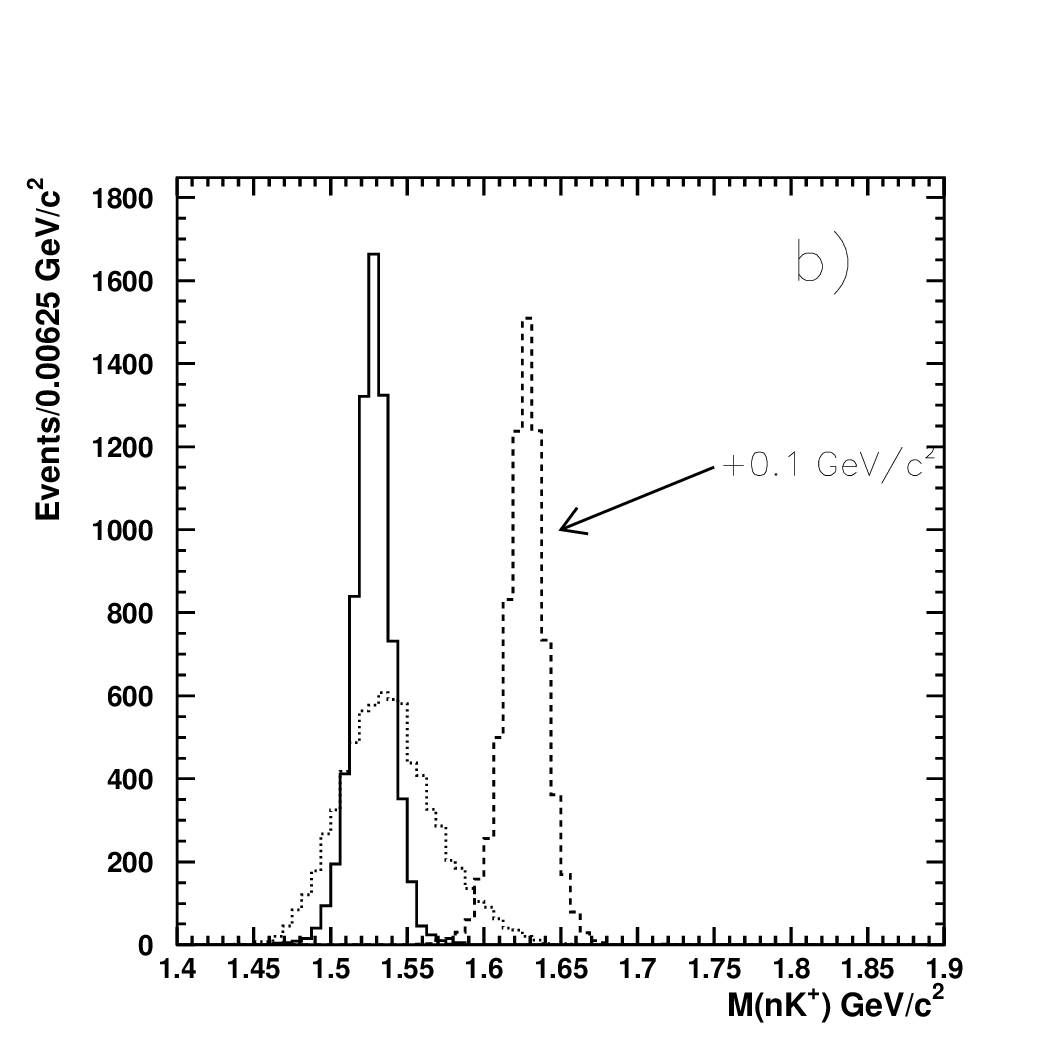}
\end{center}
\end{minipage} \\
\begin{minipage}{0.5\hsize}
\begin{center}
 \includegraphics[width=7.5cm,height=7.5cm,keepaspectratio]
{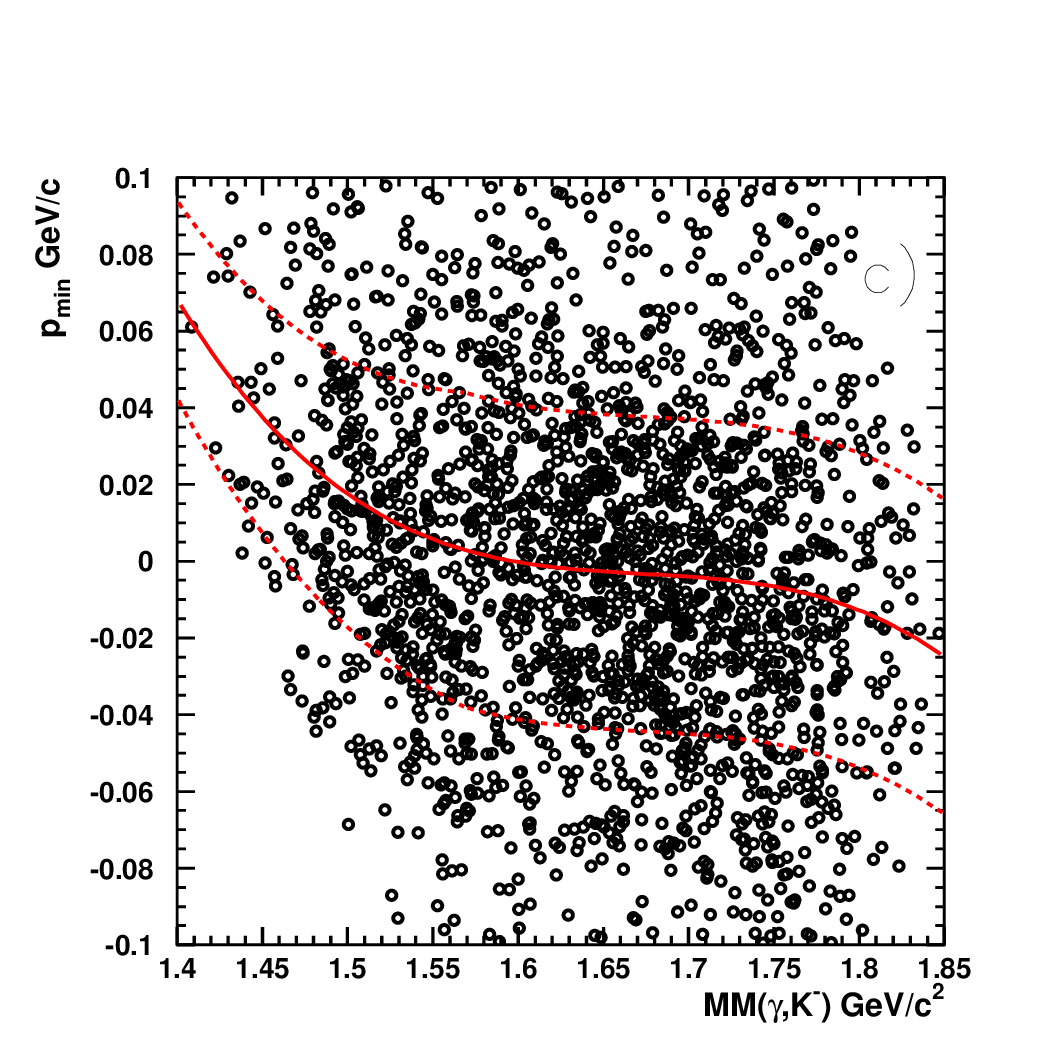}
\end{center}
\end{minipage}
\end{tabular}
\caption{ (a) $\Delta M=M(nK^+)-MM(\gamma ,K^-)$ \textit{vs.} $p_{\min } $
with a fit to a 2nd order polynomial function (solid curve).  (b)
$M(nK^+)$ (solid histogram) and $MM(\gamma,K^-)+\Delta M'$ (dashed
histogram) distributions for the signal Monte-Carlo events. The dotted
histogram is $MM(\gamma,K^-)$. (c) 2-d plot of $p_{\min}$ \textit{vs.} $MM(\gamma,K^-)$ for
$K^+K^-$ events for LD$_2$ runs. The mean and $\pm 1 \sigma$ are indicated 
by solid and dashed lines, respectively. }
\label{fig:approx}
\end{figure}

The first step of the randomized minimum momentum method (RMM) is to
approximate the mass correction ($\Delta M=M(nK^+)-MM(\gamma ,K^-))$
by a 2nd order polynomial function of $p_{\min}$ ($\equiv \Delta
M'(p_{\min})$) as shown in Fig.~\ref{fig:approx}(a). The quality of
this approximation is quite accurate, and the standard deviation of
$\Delta M-\Delta M'$ is 4-5 MeV/$c^{2 }$ in the whole mass
range. Fig.~\ref{fig:approx}(b) shows the $M(nK^+)$ and $MM(\gamma
,K^-)+\Delta M'$ distributions for the signal Monte-Carlo events. It
demonstrates that the mass correction is predominantly determined by
$p_{\min } $, and other effects such as the directions of
$\overrightarrow p _{K^+} $ and $\overrightarrow p _n $ are small.
The practical advantage of this simplified mass calculation is
separation of input arguments for the mass function into two types:
one which depends on only $E_\gamma $ and $\overrightarrow p _{K^-} $,
and the other which also depends on $\overrightarrow p _{K^+} $. The
original $M(nK^+)$ with the MMSA is a complicated function of
$E_\gamma $, $\overrightarrow p _{K^-} $, and $\overrightarrow p
_{K^+} $. In the RMM, it is approximated by a function of $MM(\gamma
,K^-)$ and $p_{\min }$.

In the next step, the most probable $M(nK^+)$ spectrum shape for a
given $MM(\gamma ,K^-)$ distribution is estimated by combining each
$MM(\gamma ,K^-)$ value with randomized $p_{\min}$ values for many
times ($10^4$ times in this analysis). In the generation, the $p_{\min}$ 
distribution is assumed to have a Gaussian shape. Because there is a weak
correlation between $p_{\min } $ and $MM(\gamma ,K^-)$ near the tails
of the $MM(\gamma ,K^-)$ distribution, the mean of the Gaussian
distribution must be varied as a function of $MM(\gamma ,K^-)$.  This
correlation is mainly caused by the difference between the kinematic
domains of $MM(\gamma ,K^-)$ and $M(nK^+)$; the former can have a
value below $m_n +m_{K^+} $, but the latter cannot. For the same
reason, the standard deviation ($\sigma$) of the $p_{\min}$
distribution must be varied near the mass threshold.  The magnitudes
of changes in the mean and $\sigma$ are small compared to a typical
$\sigma$ value of $\sim $40 MeV/$c^{2}$. Fig.~\ref{fig:approx}(c)
shows the mean and $\pm 1 \sigma$ curves as functions of $MM(\gamma
,K^-)$. The same functions are used for the estimation of the $M(pK^-)$
spectrum shape from $MM(\gamma , K^+)$.

The RMM is similar to the mixed event technique which is widely used
for estimations of combinatoric background. Both methods require
the independence of uncorrelated variables, and a common problem is
signal contamination. In the case of the RMM, the signal contamination
causes enhancement of the background level in the region of interest.

\begin{figure}[htbp]
\begin{tabular}{c c}
\begin{minipage}{0.5\hsize}
\begin{center}
 \includegraphics[width=7.5cm,height=7.5cm,keepaspectratio]
{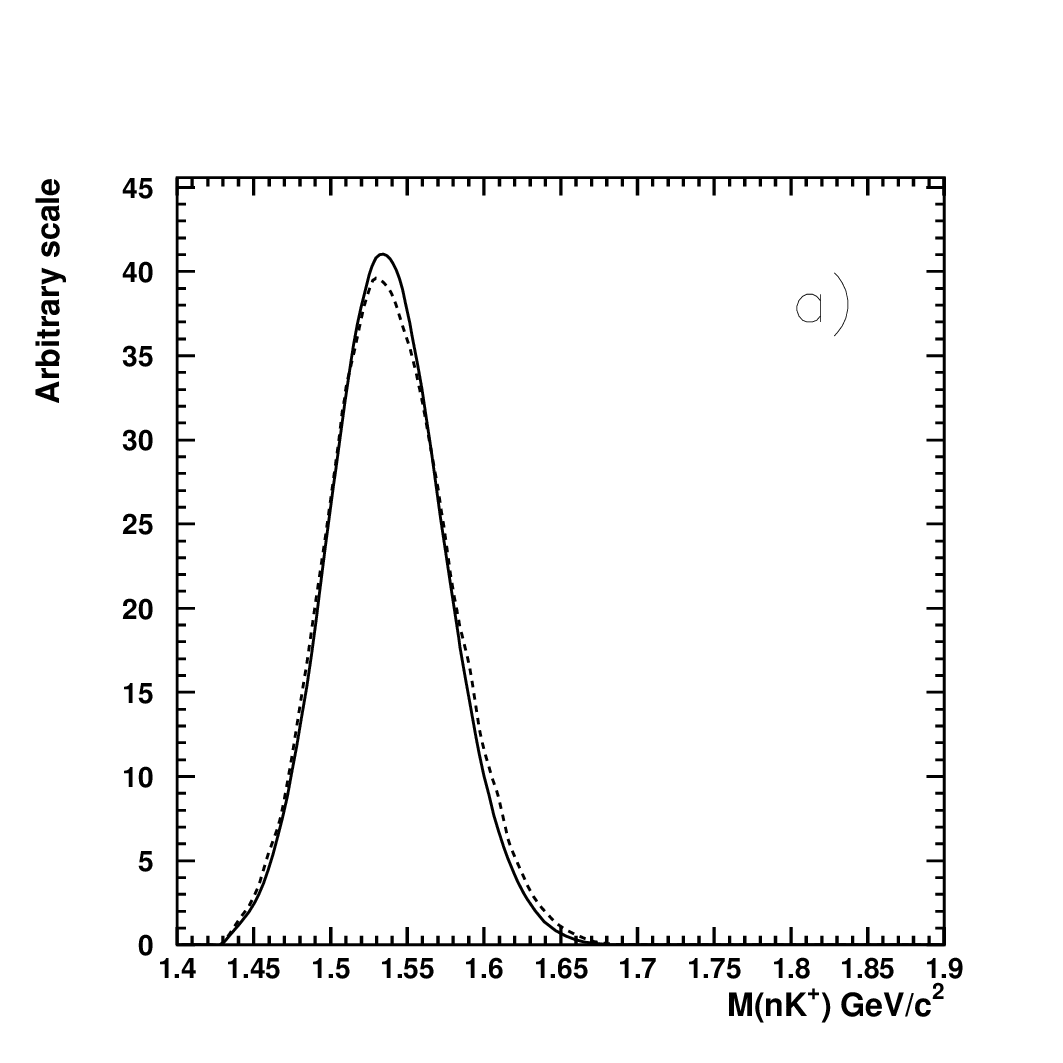}
\end{center}
\end{minipage} & %
\begin{minipage}{0.5\hsize}
\begin{center}
 \includegraphics[width=7.5cm,height=7.5cm,keepaspectratio]
{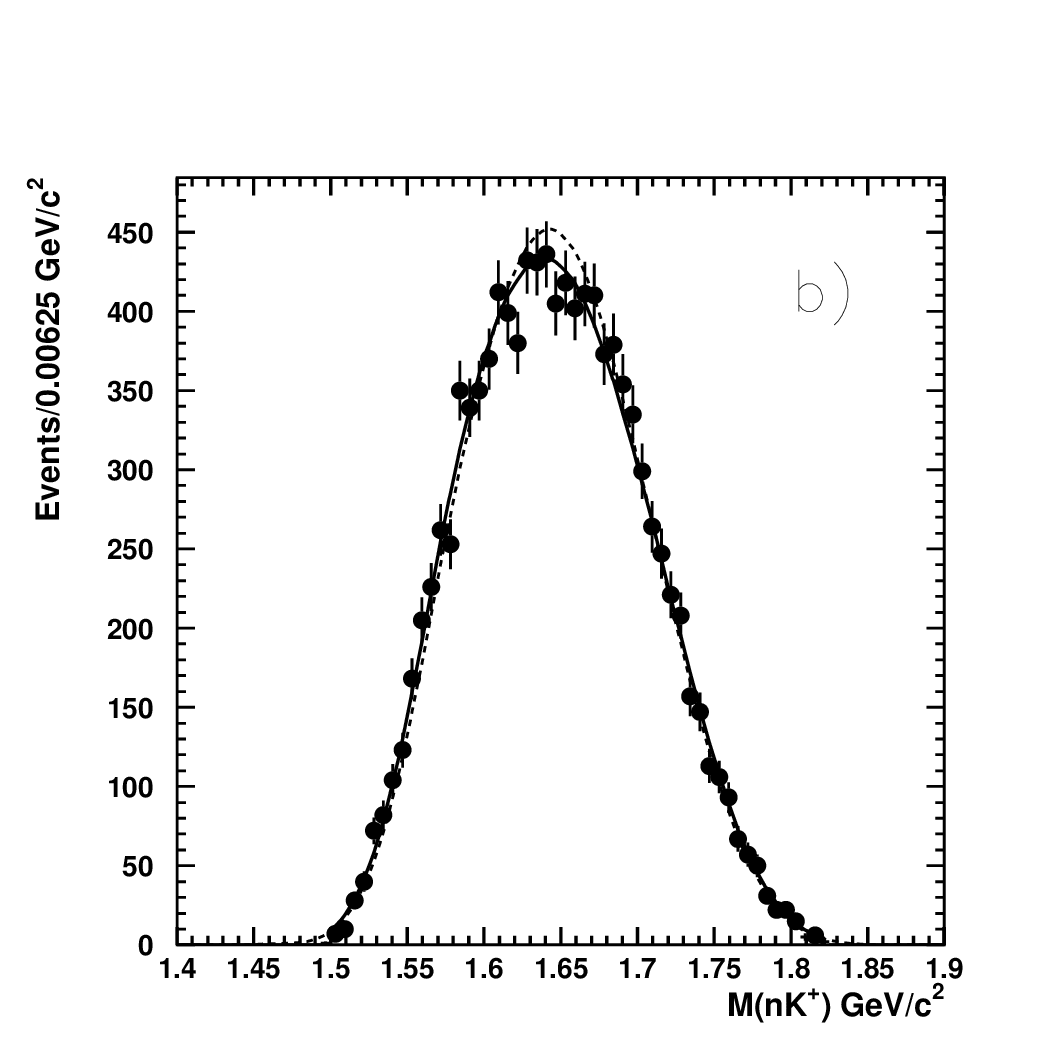} 
\end{center}
\end{minipage} \\
\begin{minipage}{0.5\hsize}
\begin{center}
 \includegraphics[width=7.5cm,height=7.5cm,keepaspectratio]
{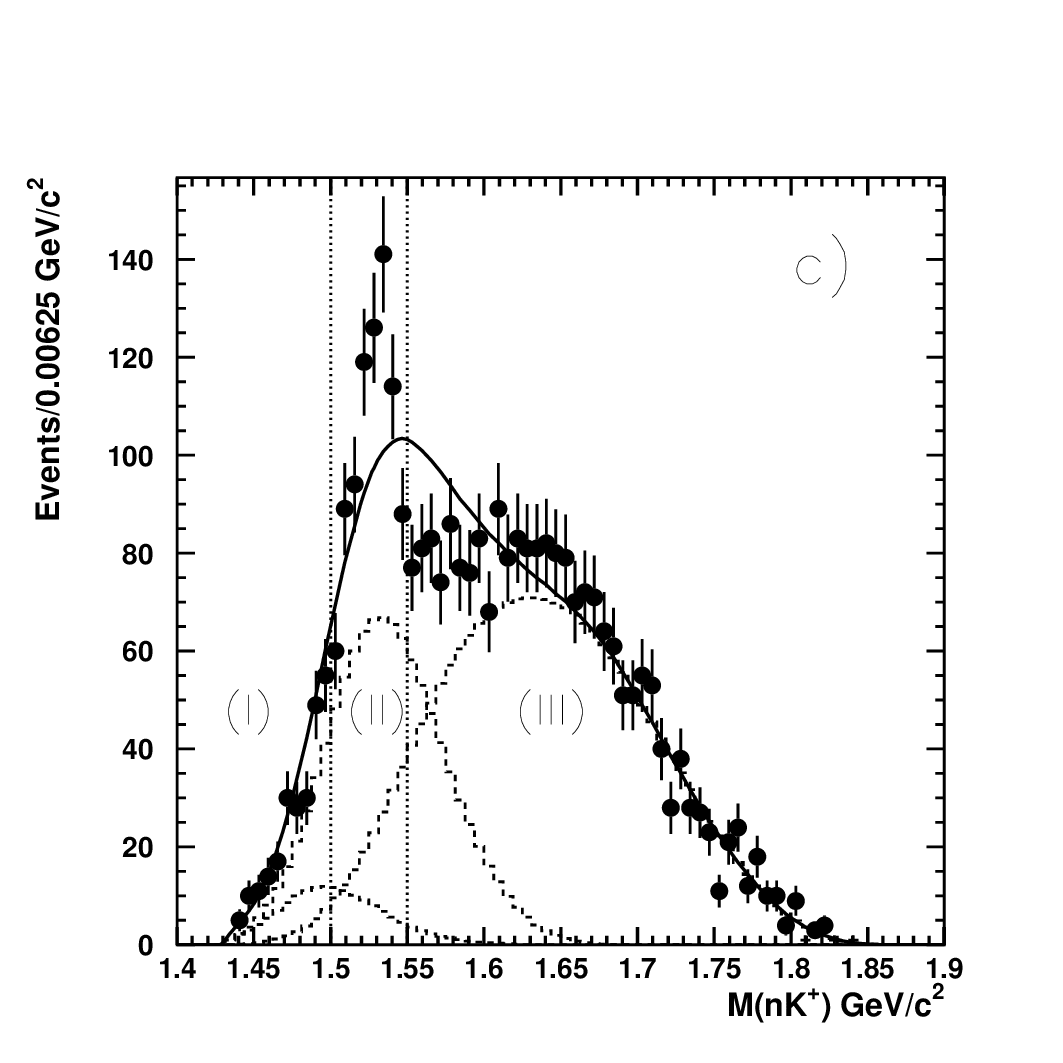}
\end{center}
\end{minipage} & %
\begin{minipage}{0.5\hsize}
\begin{center}
 \includegraphics[width=7.5cm,height=7.5cm,keepaspectratio]
{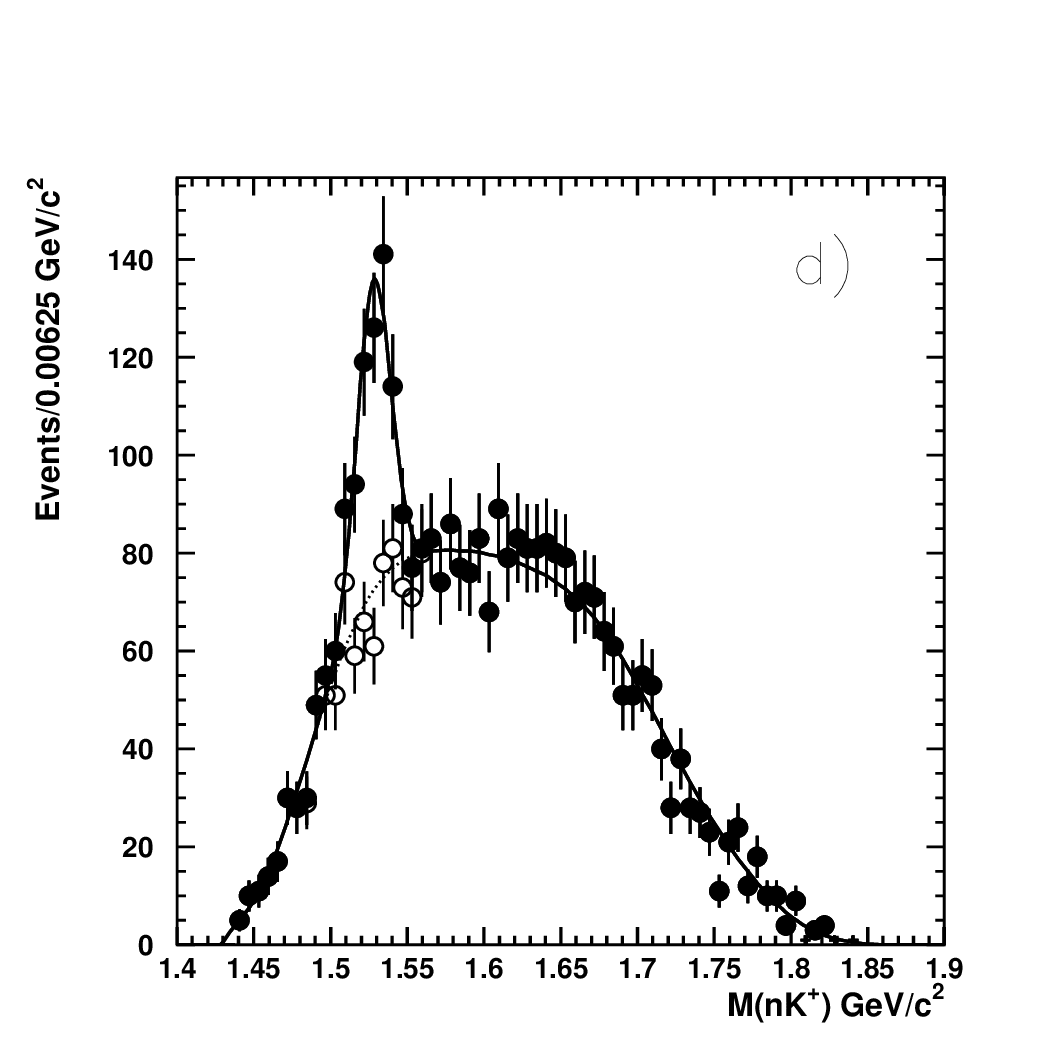} \\
\end{center}
\end{minipage}
\end{tabular}
\caption{ (a) RMM spectrum for $M(nK^+)$ distribution generated from
the signal MC events (solid line) and from non-resonant $K^+K^-$
events with 1.50 GeV/$c^2$ $< M(nK^+) <$ 1.55 GeV/$c^2$.  (b)
$M(nK^+)$ distribution for $\phi $ events and a fit to the RMM spectra
with one seed set (dashed line) and three seed sets (solid line).  (c)
$M(nK^+)$ distribution for the sum of non-resonant $K^+K^-$ MC events
and $\Theta^+$ MC events and a fit (solid curve) to a mass
distribution consisting of RMM distributions with three seed regions;
(I), (II), and (III). Contributions from each seed region is indicated
by a dashed histogram.  (d) (a) $M(nK^+)$ distribution (closed circle)
for the MC events with a fit to a distribution consisting of RMM
spectra and a Gaussian function (solid line). The dotted line is the
background contribution (the sum of the RMM spectra with fitted weight
parameters). $M(nK^+)$ distribution for non-resonant $K^+K^-$ events
(open circle). }
\label{fig:rmm}
\end{figure}

The final step of the RMM is to divide the real data events which are
used for the seeds of the event generation into several sub sets.  We
call them seed sets. In the current analysis, events are sorted
according to the value of $M(nK^+)$ or $M(pK^-)$ which is most
sensitive to the signal-to-noise ratio. The boundary for one of the
seed sets is chosen to cover a signal region or more precisely a
possible signal region. The events of the seed set for the signal
region would contain both signal events and background events. A good
feature of the RMM is that the shape of the mass distribution
generated with the signal events is very close to the shape of the
mass distribution generated with the background events in the same
seed set. Fig.~\ref{fig:rmm}(a) shows the RMM spectra generated with
the $\Theta^+$ MC events and with the non-resonant $K^+K^-$ MC events
both in the region of 1.50 GeV/$c^2$ $<M(nK^+)<$ 1.55 GeV/$c^2$. The
difference in the spectrum shape is small.

In the shape analysis in the following sections, the background
spectrum is represented as a sum of several RMM spectra. If one of the
seed sets is contaminated by signal events, the effect can be absorbed
by reduction of the weight parameter in the summation of the RMM
spectra.  By using several RMM spectra for the shape analysis, some of
the global inconsistencies caused by fluctuations in the seed
distributions and incomplete treatment of the correlation between
$p_{\min }$ and $MM(\gamma ,K^\pm )$ can also be compensated by small
changes of the weight parameters.

The significance of a signal contribution is calculated from the
difference in log likelihood between fits with and without the signal 
contribution represented by a Gaussian function. Since the width
is fixed to the value estimated by a Monte-Carlo simulation, 
the change in the number of the degrees of freedom is 2, which is taken
into account for the significance calculation. 

The $M(nK^+)$ distribution for $\phi $ events selected with
requirements of 1.01 GeV/$c^{2} <M(K^+K^-)<$1.03 GeV/$c^{2}$ and 2.0
GeV $<E_\gamma ^{eff} <$ 2.5 GeV is shown in Fig.~\ref{fig:rmm}(b). A
fit to a spectrum generated with the RMM using all selected events
with an equal weight is indicated by the dashed line. The solid line
shows the fit results with three RMM spectra, for which the selected
$\phi$ events are divided into the three seed sets according to the
conditions: $M(nK^+)<$1.5 GeV/$c^{2}$, 1.5 GeV/$c^{2} <M(nK^+)<$ 1.6
GeV/$c^{2}$, or $M(nK^+)>$ 1.6 GeV/$c^{2}$. The log likelihood
(-2ln$L)$ for the single RMM spectrum fit is 65.5 for the number of
the degrees of freedom ($ndf$) equal to 51. It becomes to 50.4 for the
fit with three RMM spectra for $\Delta ndf = 2$. Since the spread of
a RMM spectrum from the seed set of 1 MeV/$c^2$ width is larger than
30 MeV/$c^2$, further increasing the segmentation for the seed sets
does not improve the -2ln$L$ value more than $\Delta ndf$.

The $M(nK^+)$ distribution for the sum of 3000 non-resonant $K^+K^-$
MC events and 300 $\Theta^+$ MC events is fitted to a mass
distribution consisting of three RMM distributions with seed regions
of (I) $M(nK^+)<$ 1.50 GeV/$c^{2}$ , (II) 1.50 GeV/$c^{2} <M(nK^+)<$
1.55 GeV/$c^{2}$, and (III) $M(nK^+)>$ 1.55 GeV/$c^{2}$.  The best
fit, which is indicated by a solid curve in Fig.~\ref{fig:rmm}(c), is
obtained with the weight parameters of 0.651, 1.245, and 0.949 for the
contrbutions from region (I), (II), and (III), respectively.  The
-2ln$L$ value for the fit is 114.6 for $ndf$ = 61. The -2ln$L$ value
is improved to 58.4 for $ndf$ = 59 by including a Gaussian function
with a fixed width of 11 MeV/c$^{2}$ to represent the $\Theta^+$
contribution as shown in Fig.~\ref{fig:rmm}(d).  The statistical
significance of the peak is calculated to be 7.2 $\sigma$ for $\Delta$
(-2ln$L$)/$\Delta ndf$ = 56.2/2.  The weight parameters become 1.14,
0.648, and 0.993 for the contrbutions from region (I), (II), and
(III), respectively.  The sum of the RMM spectra with the fitted
weight parameters, which is indicated by a dotted curve in
Fig.~\ref{fig:rmm}(d), reproduces well the original mass distribution
(open circles) for the non-resonant $K^+K^-$ MC events.  The signal
yield estimated from the fit is 279 $\pm$ 36 events, which is
consistent with the number of $\Theta^+$ MC events of 300.

\section{ANALYSIS OF THE $\gamma p\to K^+\Lambda (1520)$ REACTION}
\begin{figure}[htbp]
\begin{tabular}{c c}
\begin{minipage}{0.5\hsize}
\begin{center}
 \includegraphics[width=7.5cm,height=7.5cm,keepaspectratio]
{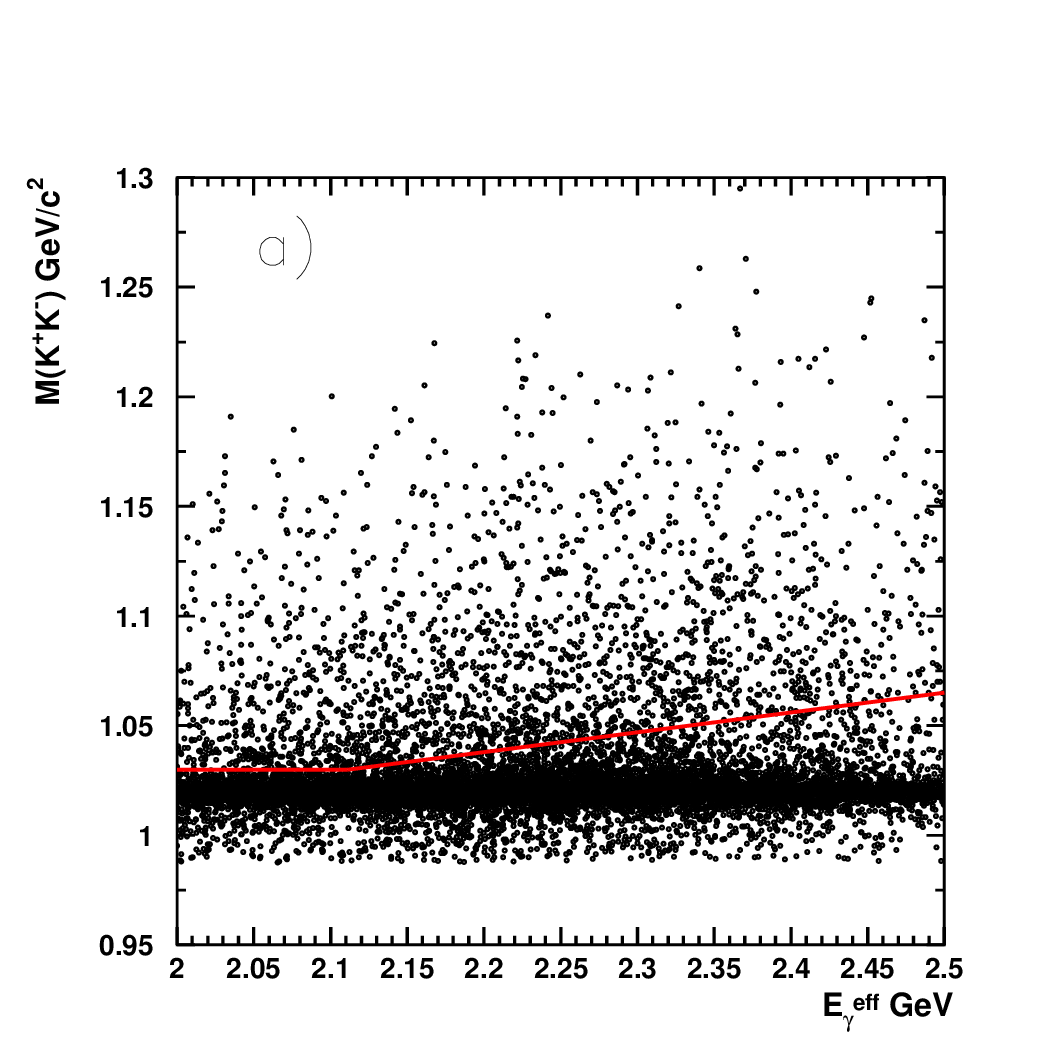}
\end{center}
\end{minipage} & %
\begin{minipage}{0.5\hsize}
\begin{center}
 \includegraphics[width=7.5cm,height=7.5cm,keepaspectratio]
{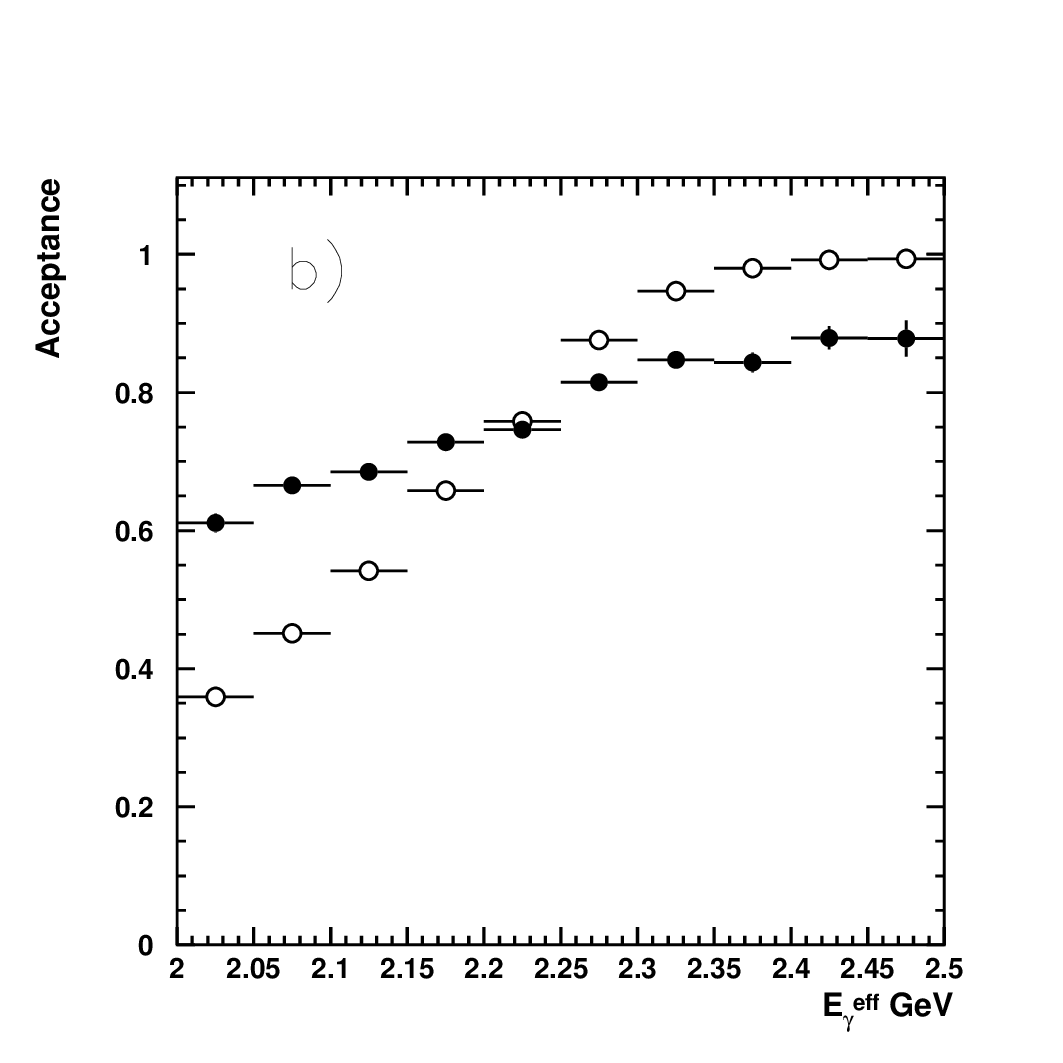}
\end{center}
\end{minipage} \\
\begin{minipage}{0.5\hsize}
\begin{center}
 \includegraphics[width=7.5cm,height=7.5cm,keepaspectratio]
{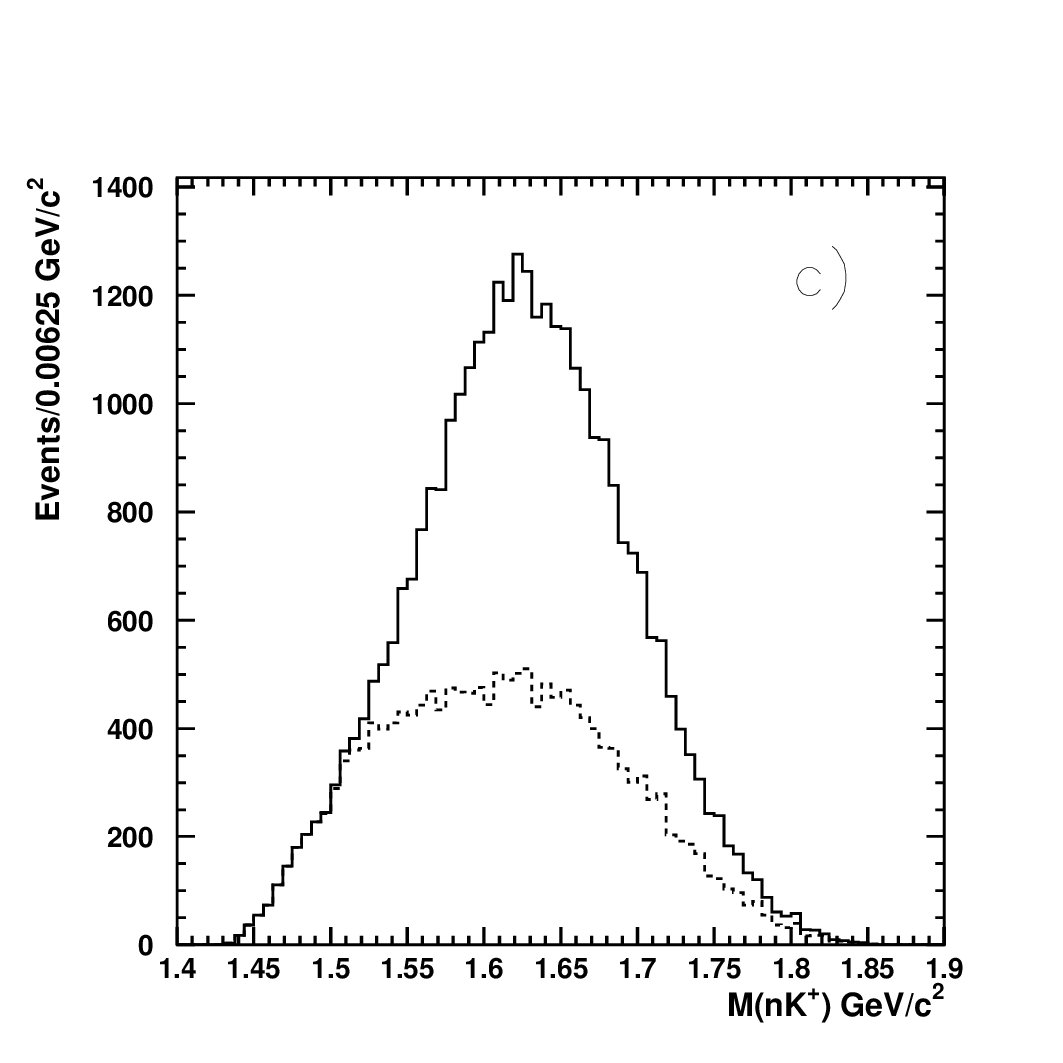}
\end{center}
\end{minipage} 
\end{tabular}
\caption{ (a) 2-d plot of $M(K^+K^-)$ \textit{vs.} $E_\gamma^{eff}$ and the cut
boundary of the $\phi$ exclusion cut (solid line).  (b) acceptances of
the $\phi$ exclusions cuts: the energy dependent cut (closed circle)
and the constant cut of $M_{KK} > 1.04$ GeV/$c^2$ (open circle).  (c)
$M(nK^+)$ distributions for non-resonant $K^+K^-$ events before the
$\phi$ exclusion cut (solid histogram) and after the cut (dashed
histogram).  }
\label{fig:phicut}
\end{figure}

The dominant contribution in the selected $K^+K^-$ events is due to $\phi
$ decays, which are rejected by the combined requirements of
$M(K^+K^-)>1.03$ GeV/$c^{2}$ and $M(K^+K^-)>1.02+0.09 \times (E_\gamma
^{eff}$ - 2.0). The cut boundary is shown
as solid lines in Fig.~\ref{fig:phicut}(a). The energy dependent 
cut condition makes the signal acceptance more uniform than the constant
cut condition as shown in Fig.~\ref{fig:phicut}(b).
The $M(nK^+)$ distributions for non-resonant $K^+K^-$ MC events
before and after the $\phi $ exclusion cut are shown in
Fig.~\ref{fig:phicut}(c). The $\phi $ exclusion cut distorts the mass
spectrum because the acceptance is high near the threshold where
the momenta of the $K^+$ and $K^-$ are highly asymmetric. However, the mass
dependence of the acceptance is not strong, and consequently the cut
does not create a narrow peak. Note that the mass spectrum near the 
threshold is not affected by the $\phi$ exclusion cut. This is because 
the momenta of two kaons for events in the threshold region are highly
asymmetric, which results in a high $K^+K^-$ mass. 

\begin{figure}[htbp]
\begin{tabular}{c c}
\begin{minipage}{0.5\hsize}
\begin{center}
 \includegraphics[width=7.5cm,height=7.5cm,keepaspectratio]
{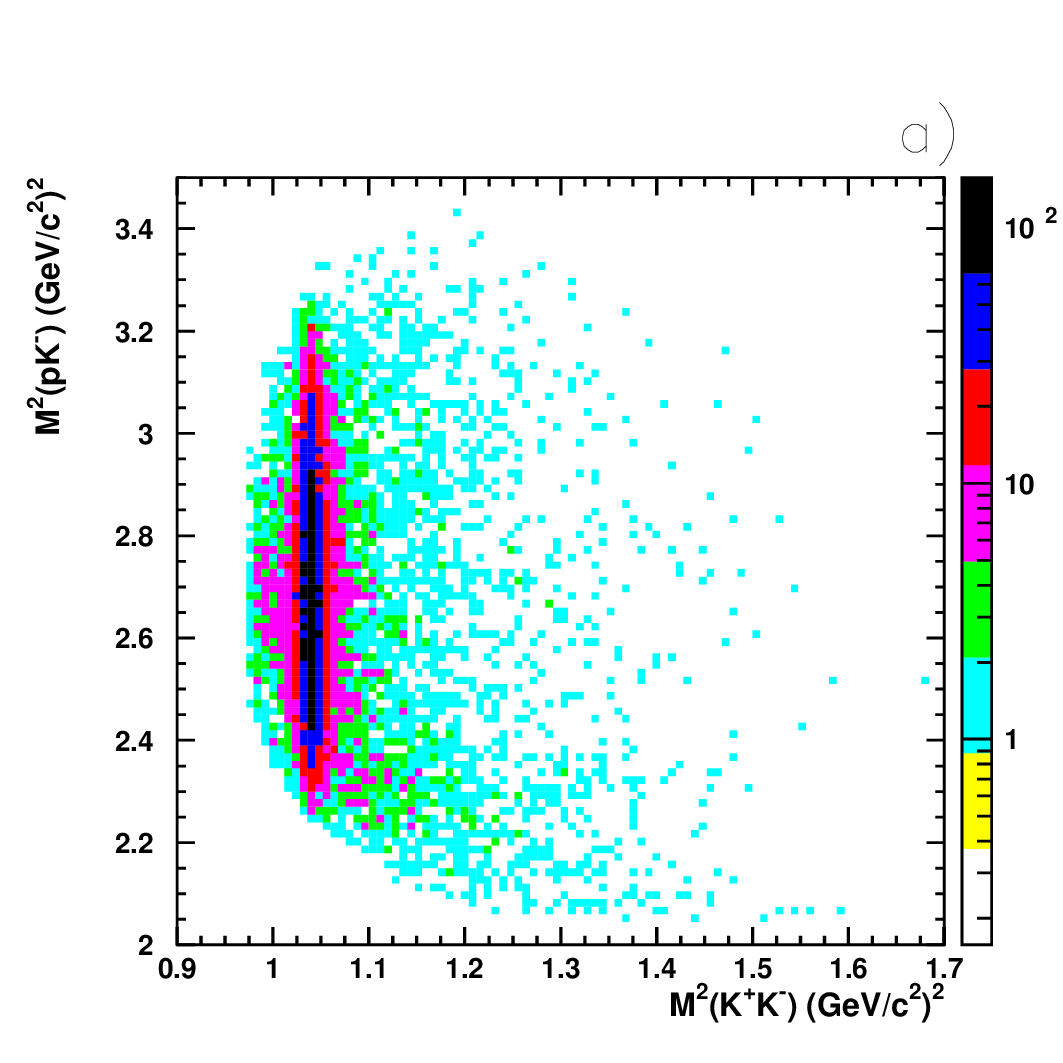}
\end{center}
\end{minipage} & %
\begin{minipage}{0.5\hsize}
\begin{center}
 \includegraphics[width=7.5cm,height=7.5cm,keepaspectratio]
{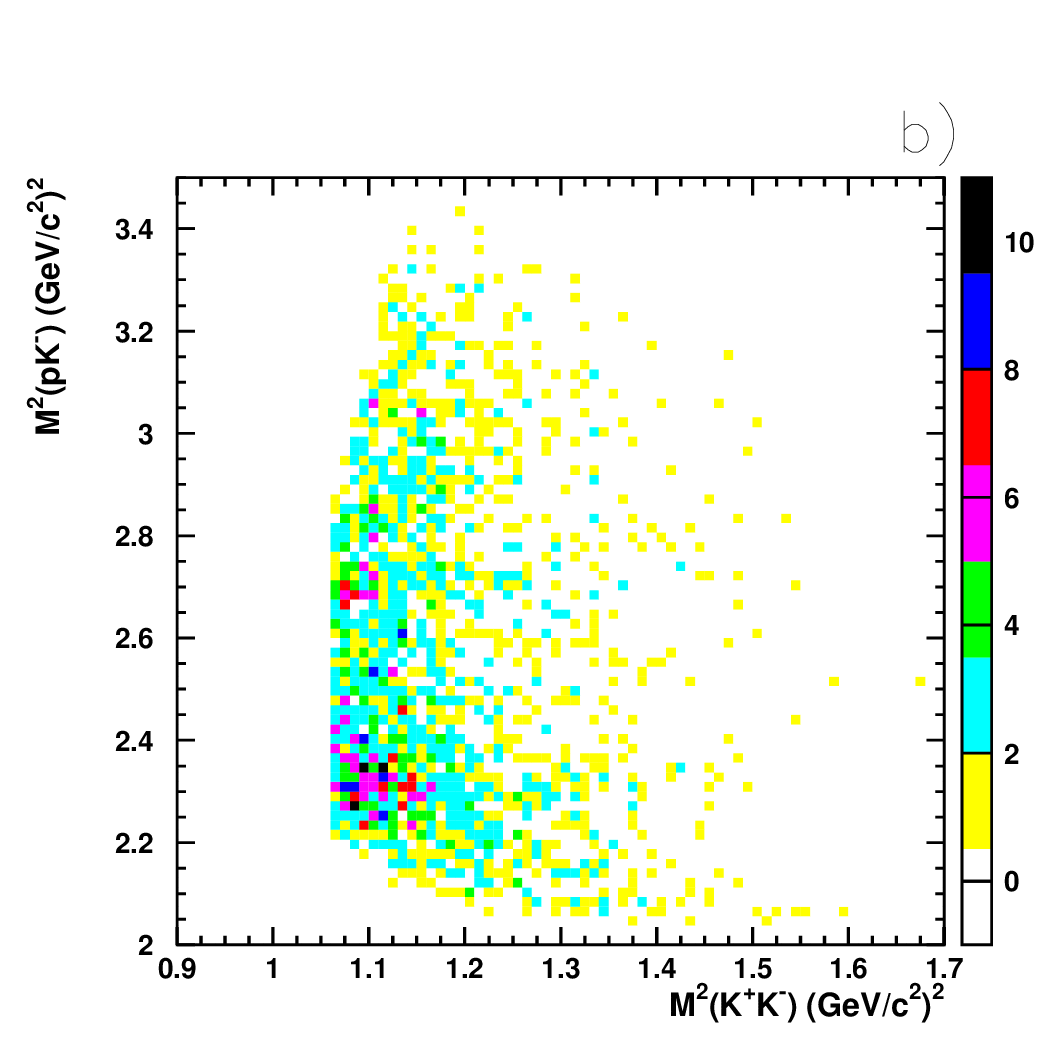}
\end{center}
\end{minipage}
\end{tabular}
\caption{
The Dalitz plots of $M^2(pK^-)$ \textit{vs.} $M^2(K^+K^-)$
before (left) and after (right) the $\phi $ exclusion cut.
}
\label{fig:kkpk}
\end{figure}

A total of 2078 events passed the $\phi $ exclusion cut, and the
MMSA is applied to obtain $M(pK^-)$ by assuming the spectator is a
neutron. The Dalitz plots ($M^2(pK^-)$ \textit{vs.} $M^2(K^+K^-))$
before and after the $\phi $ exclusion cut are shown in
Fig.~\ref{fig:kkpk}. The $\Lambda $(1520) yield at $M^2(pK^-)\sim
2.3$ GeV/$c^{2}$ is higher in the lower $M^2(K^+K^-)$ region
due to the LEPS detector acceptance.  However,
the events are not concentrated near the cut boundary.

\begin{figure}[htbp]
\begin{tabular}{c c}
\begin{minipage}{0.5\hsize}
\begin{center}
 \includegraphics[width=7.5cm,height=7.5cm,keepaspectratio]
{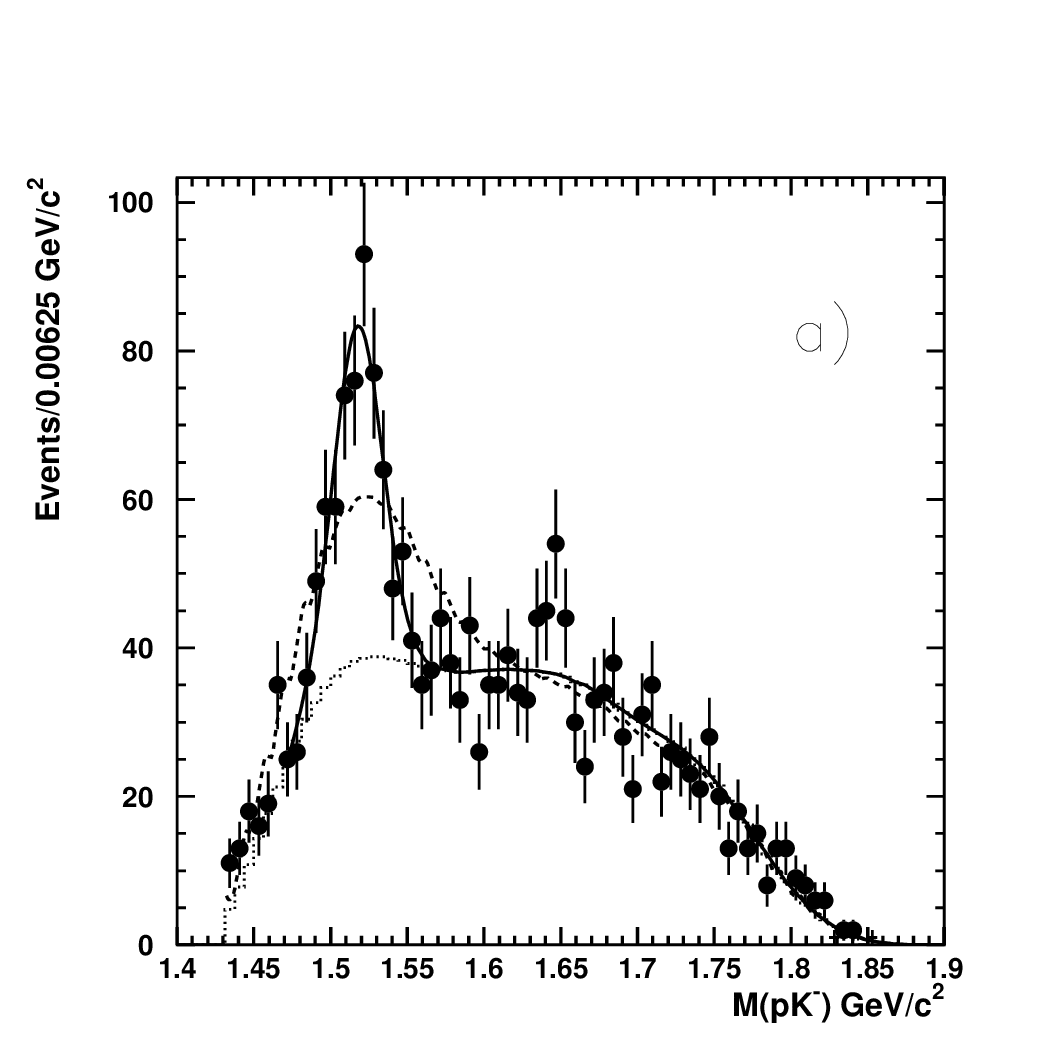}
\end{center}
\end{minipage} & %
\begin{minipage}{0.5\hsize}
\begin{center}
 \includegraphics[width=7.5cm,height=7.5cm,keepaspectratio]
{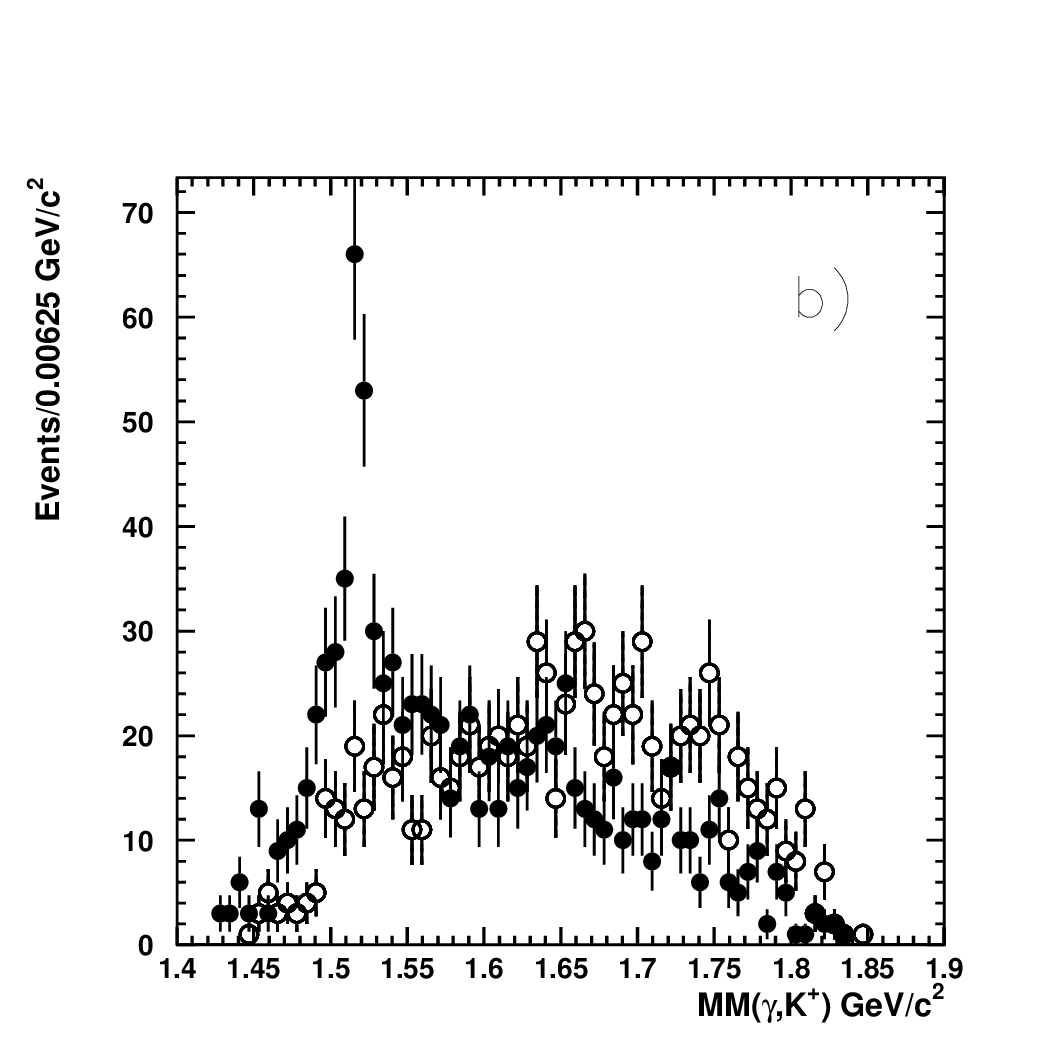}
\end{center}
\end{minipage}
\end{tabular}
\caption{ 
(a) $M(pK^-)$ distribution with a fit to the RMM background
spectrum only (dashed line) and with a Gaussian function (solid
line). The dotted line is the background.  
(b) $MM(\gamma, K^+)$ (closed circle) and $MM(\gamma, K^-)$ distributions for
the LH$_2$ runs.}
\label{fig:pk}
\end{figure}

Fig.~\ref{fig:pk}(a) shows the $M(pK^-)$ distribution. For the shape
analysis, three RMM spectra are generated by setting the seed
boundaries at 1.48 GeV/$c^{2}$ and 1.56 GeV/$c^{2}$ in $M(pK^-)$.  A
fit to the RMM spectra gives a -2ln$L$ value of 110.2 for
$ndf$=58. The -2ln$L$ value is improved to 55.1 by including a
Gaussian function with a fixed width of 16 MeV/c$^{2}$ as the $\Lambda
$(1520) contribution. The $\Delta $(-2ln$L)$ of 55.1 for $\Delta
ndf$=2 corresponds to a statistical significance of
7.1$\sigma $. The signal yield is determined to be 289$\pm $38 events
from the fit. The fit result with the $\Lambda $(1520) contribution is
represented by the solid curve. The dotted line is the sum of the RMM
spectra, which represents the background. The dashed line shows a
fitting result without the $\Lambda $(1520) contribution.

The data points of the $MM(\gamma ,K^+)$ and $MM(\gamma ,K^-)$
distributions for the LH$_{2}$ runs are shown in Fig.~\ref{fig:pk}(b)
as closed circles and open circles, respectively. The $\Lambda $(1520)
peak becomes narrow because of no Fermi motion effect. No significant
peak structure is observed in the $MM(\gamma ,K^-)$ distribution. The
mass distribution in the region of 1.47 GeV/$c^{2} <MM(\gamma
,K^-)<$ 1.65 GeV/$c^{2}$ is fitted to a second-order polynomial. The
result gives -2ln$L$=32.3 for $ndf$=27. An excess of the
$MM(\gamma ,K^+)$ yield over the $MM(\gamma ,K^-)$ yield is seen near
the $NK$ mass threshold. The excess of the $MM(\gamma ,K^-)$ yield
in the high mass regions is due to the reflections of the $\Lambda
$(1520) events. The $MM(\gamma ,K^+)$ distribution is fitted to a
Gaussian function plus a second-order polynomial, and the $\Lambda
$(1520) yield is estimated to be 143$\pm $17 from the fit. The
LD$_{2}$/LH$_{2}$ ratio of the $\Lambda $(1520) yield is 2.02$\pm
$0.36. This is consistent with the estimated ratio of 1.93 from the
integrated numbers of incident photons and the target proton
densities.

\section{ANALYSIS OF THE $\gamma n\to K^-\Theta ^+$ REACTION}
For the analysis of the $\gamma n\to K^-\Theta ^+$ reaction, events
with a 3rd charged track in addition to $K^+$ and $K^-$ tracks are
removed. This condition changes the total number of events from 2078
to 1967. Most of the removed events are due to $\gamma p\to K^+K^-p$
reactions with a neutron as a spectator, for which the LEPS detector
has a finite acceptance to detect all the charged particles in the
final state.

\begin{figure}[htbp]
\begin{tabular}{c c}
\begin{minipage}{0.5\hsize}
\begin{center}
 \includegraphics[width=7.5cm,height=7.5cm,keepaspectratio]
{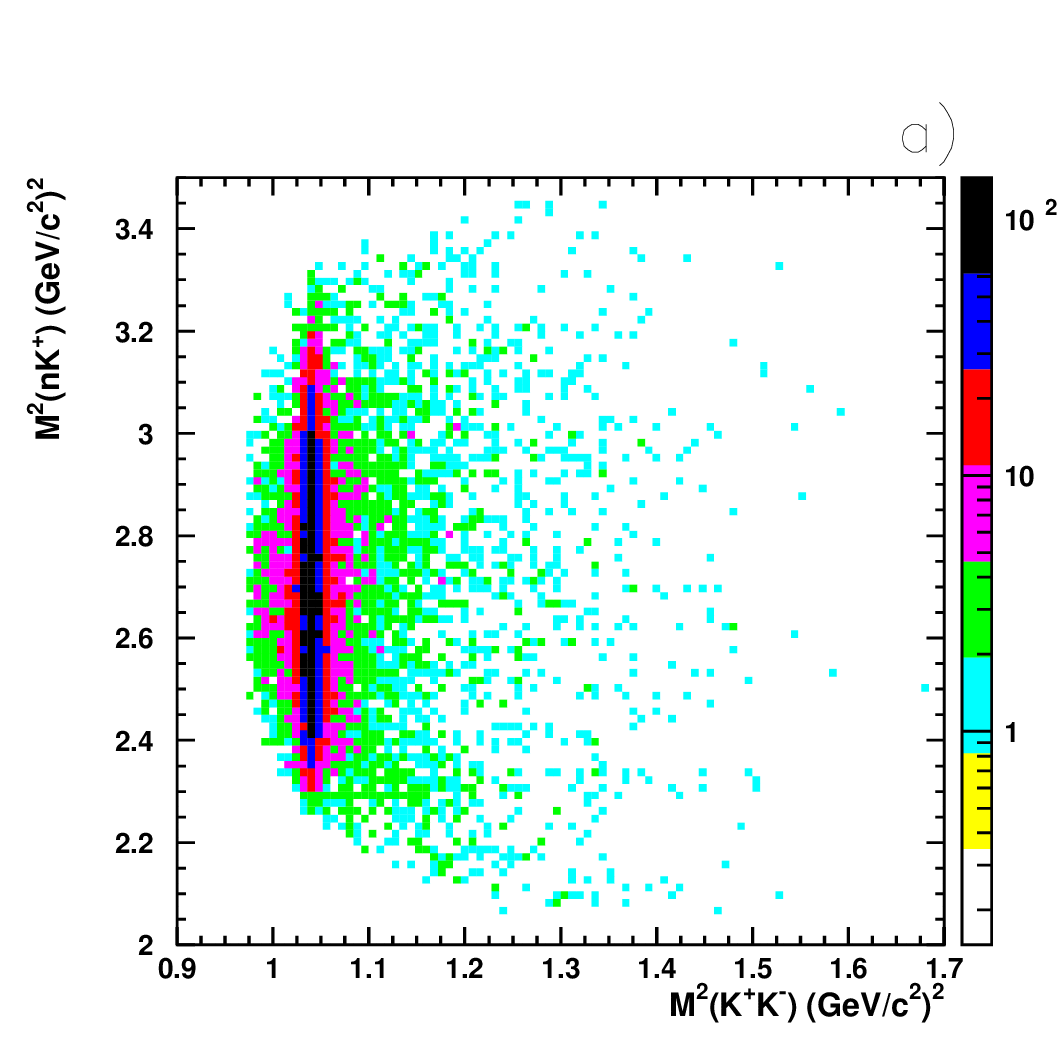}
\end{center}
\end{minipage} & %
\begin{minipage}{0.5\hsize}
\begin{center}
 \includegraphics[width=7.5cm,height=7.5cm,keepaspectratio]
{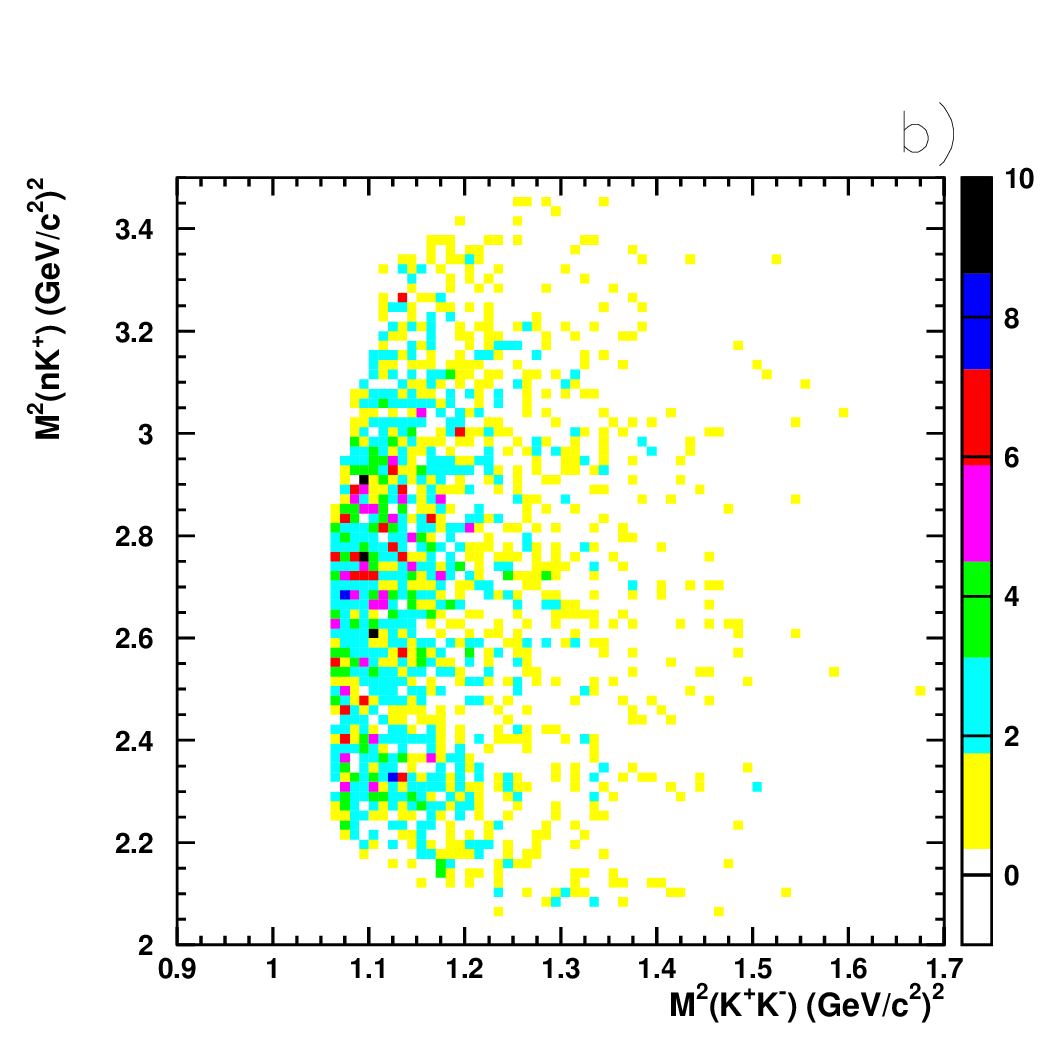}
\end{center}
\end{minipage}
\end{tabular}
\caption{
The Dalitz plots of $M^2(nK^+)$ \textit{vs.} $M^2(K^+K^-)$
before (left) and after (right) the $\phi $ exclusion cut.
}
\label{fig:phicut2}
\end{figure}

\begin{figure}[htbp]
\begin{tabular}{c c}
\begin{minipage}{0.5\hsize}
\begin{center}
 \includegraphics[width=7.5cm,height=7.5cm,keepaspectratio]
{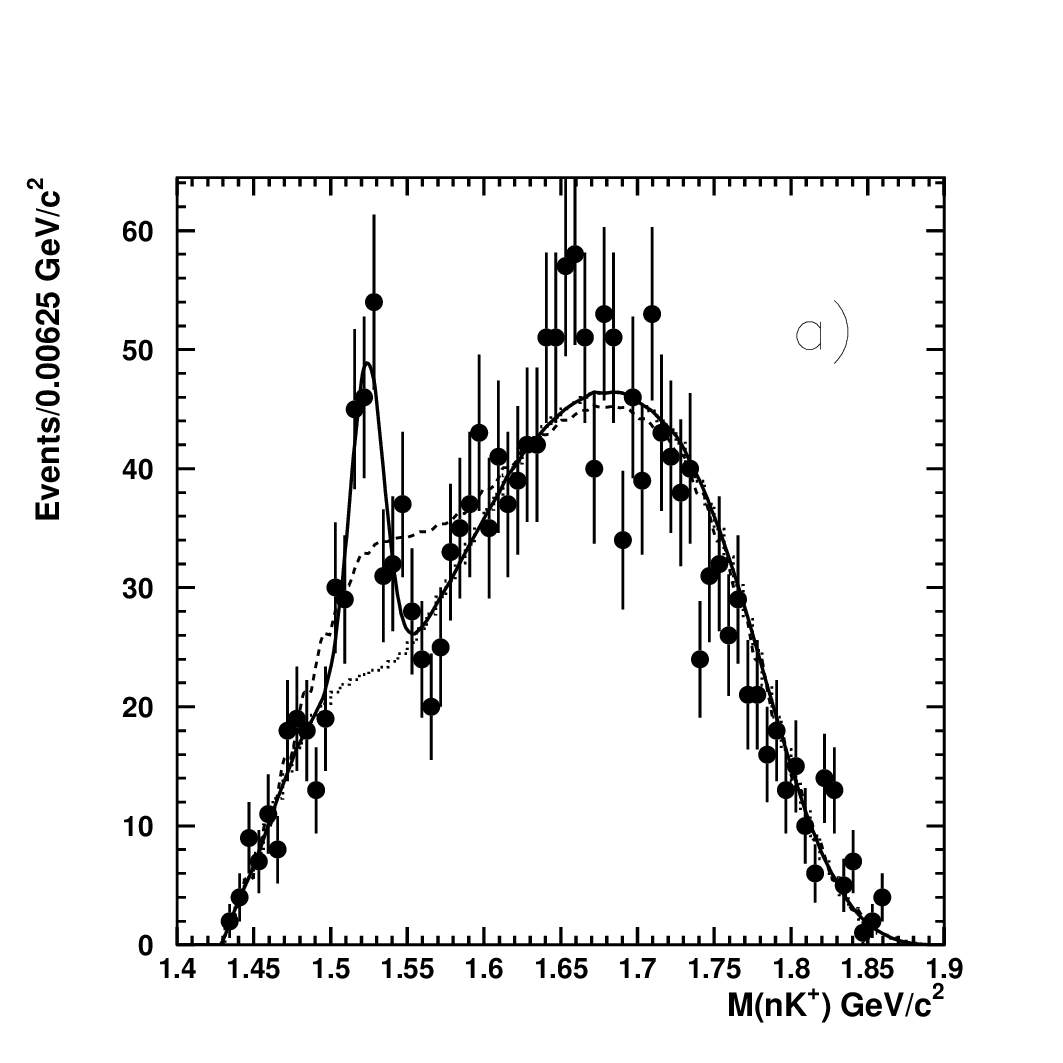}
\end{center}
\end{minipage} & %
\begin{minipage}{0.5\hsize}
\begin{center}
 \includegraphics[width=7.5cm,height=7.5cm,keepaspectratio]
{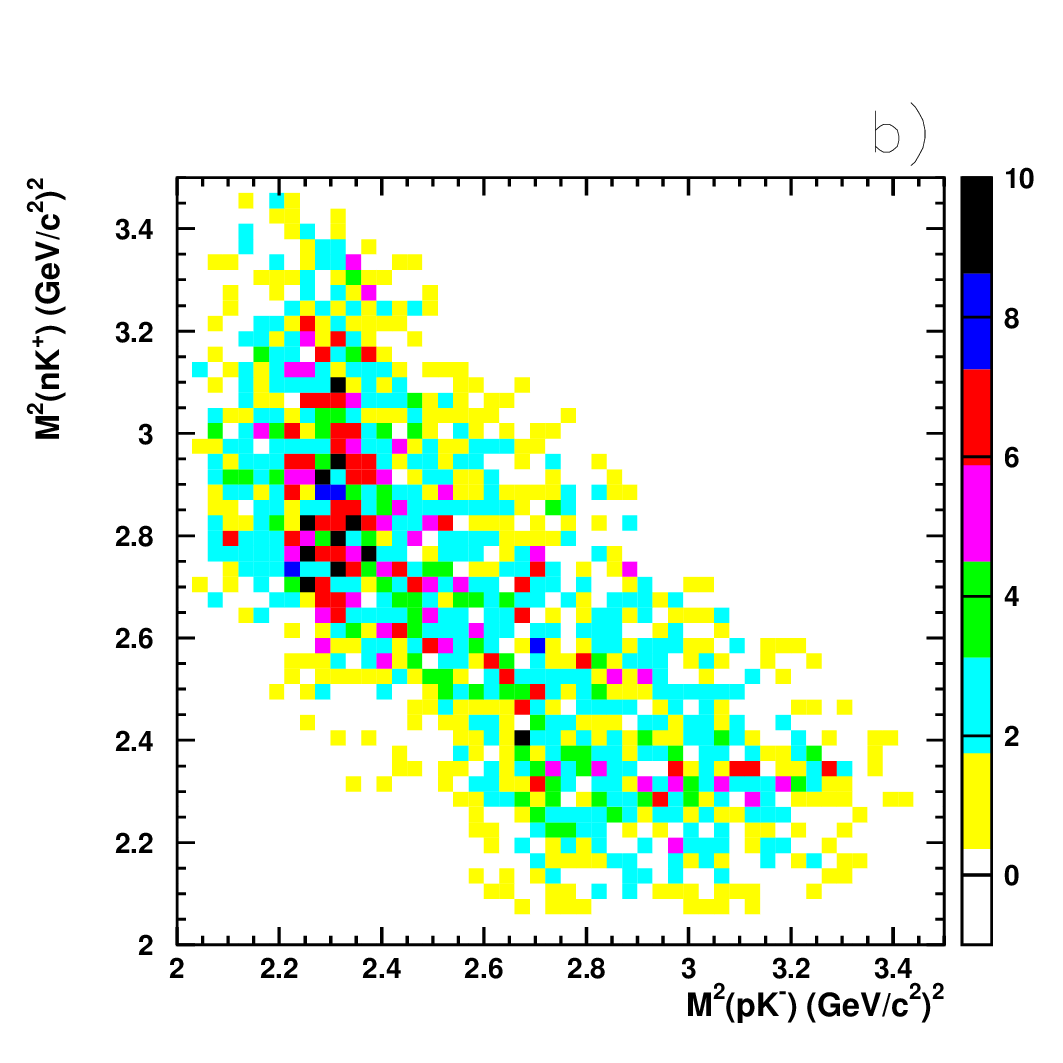}
\end{center}
\end{minipage}
\end{tabular}
\caption{
(a) $M(nK^+)$ distribution with a fit to the RMM background
spectrum only (dashed line) and with a Gaussian function (solid
line). The dotted line is the background.  
(b) Dalitz plot of $M^2(nK^+)$ \textit{vs.} $M^2(pK^-)$.
}
\label{fig:nk}
\end{figure}

Dalitz plots of $M^2(nK^+)$ \textit{vs.}  $M^2(K^+K^-)$ before and
after the $\phi $ exclusion cut are shown in
Fig.~\ref{fig:phicut2}. No concentration of events near the cut
boundary is seen. The $M(nK^+)$ distribution for the final candidate
events is shown in Fig.~\ref{fig:nk}(a). There is a narrow peak
structure near 1.52-1.53 GeV/$c^{2}$. The distribution is fitted to a
mass distribution consisting of three RMM distributions with seed
regions of $M(nK^+)<$ 1.50 GeV/$c^{2}$ , 1.50 GeV/$c^{2} <M(nK^+)<$
1.55 GeV/$c^{2}$, and $M(nK^+)>$ 1.55 GeV/$c^{2}$. The -2ln$L$ value
of the fit changes from 104.7 (for $ndf$=66) to 73.64 (for
$ndf$=64) by including a Gaussian function with the estimated
width of 11 MeV/$c^{2}$ to represent the $\Theta ^{+}$ signals.  The
statistical significance of the signal estimated from $\Delta
$(-2ln$L$) is 5.2$\sigma $. The peak position for the best fit is
$1.524\pm 0.002+0.003$ GeV/$c^{2}$, where the systematic shift of the
peak position by +3 MeV/$c^{2}$ due to the MMSA and the $\phi $
exclusion cut is given as a systematic uncertainty. The signal yield
is estimated to be 116 $\pm$ 21 events from the fitted peak height and
its uncertainty.  The detector acceptance has been calculated by
assuming the isotropic production of the $\Theta^+$ in the $\gamma n$
center-of-mass system, and the differential cross-section for the
$\gamma n \to K^- \Theta^+$ reaction is estimated to be $12 \pm 2$
nb/sr in the LEPS angular range.

There is a dip near 1.56 GeV/$c^2$ even with the $\Theta^+$
contribution.  However, with the current limited statistics, it is not
clear if the dip is due to fluctuations or due to some interference
effects. Since we assume the branching ratio B($\Theta^+ \to K^+ n$)=0.5
in the calculation of the differential cross-section, possible
interference effects between the signal and background amplitudes
could result in a change of the estimated value.

A fit of the $M(nK^+)$ distribution to the mass distribution using a
Gaussian function with a free width parameter has been carried out,
and the best fit is obtained with a width of 12.7 $\pm $2.8
MeV/$c^{2}$, which is consistent with the estimated width of 11
MeV/$c^{2}$.

\begin{figure}[htbp]
\begin{tabular}{c c}
\begin{minipage}{0.5\hsize}
\begin{center}
 \includegraphics[width=7.5cm,height=7.5cm,keepaspectratio]
{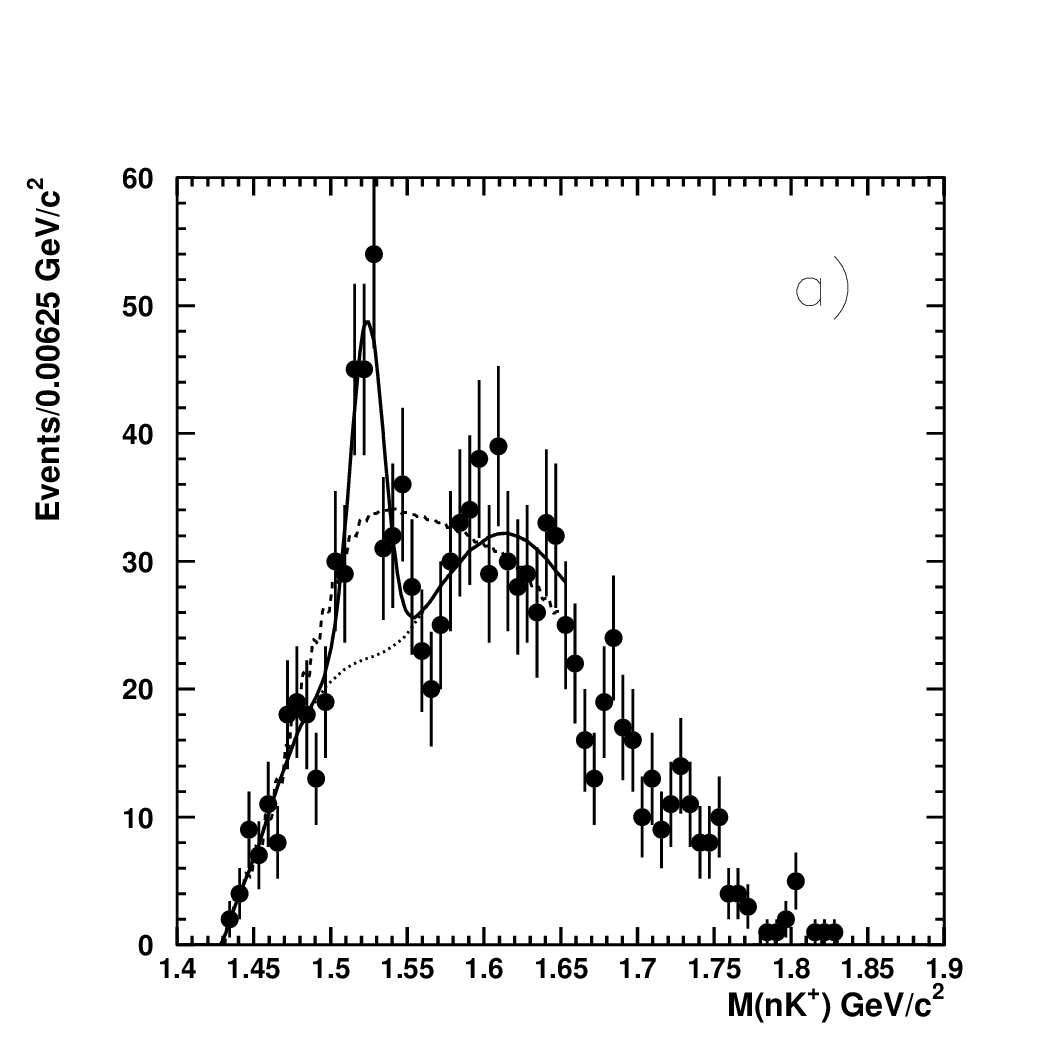}
\end{center}
\end{minipage} & %
\begin{minipage}{0.5\hsize}
\begin{center}
 \includegraphics[width=7.5cm,height=7.5cm,keepaspectratio]
{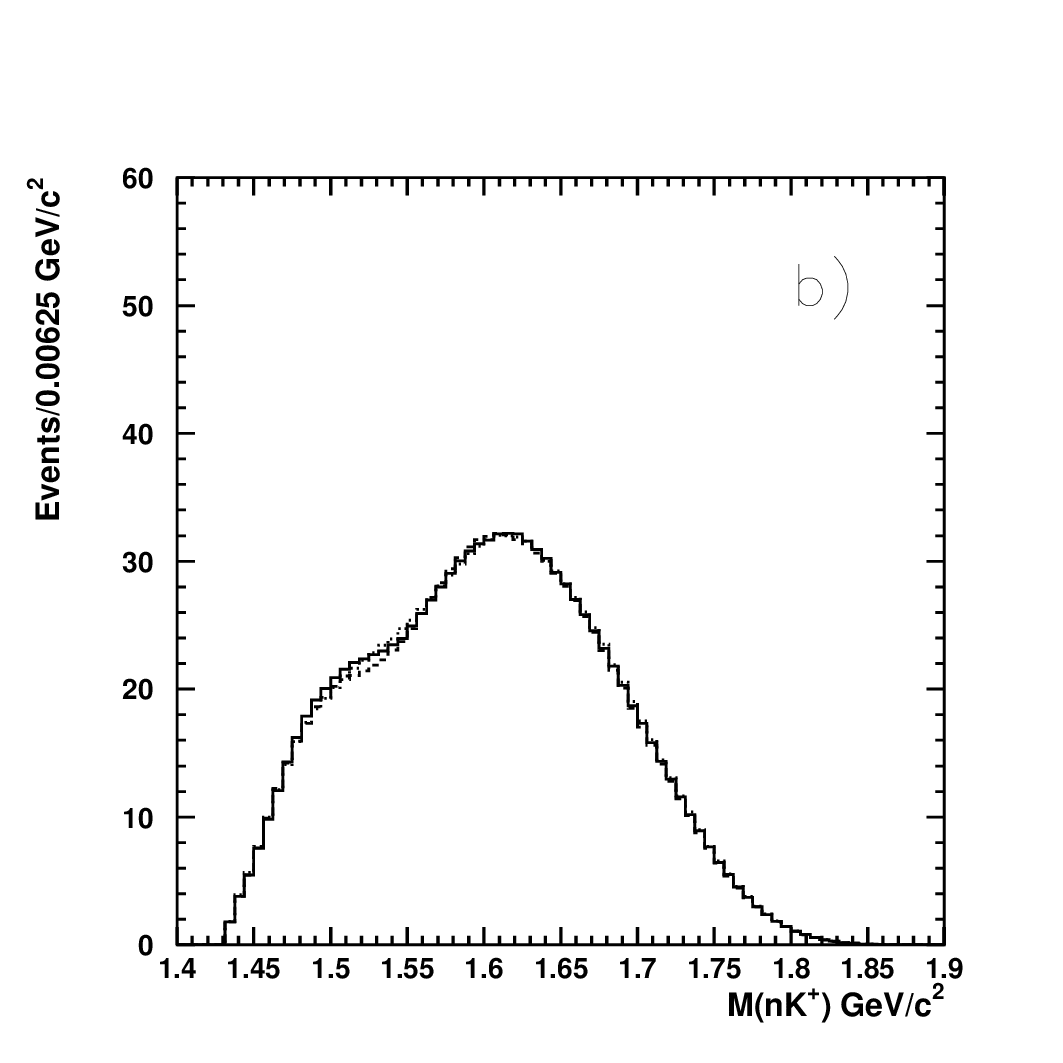}
\end{center}
\end{minipage}
\end{tabular}
\caption{
(a) $M(nK^+)$ distribution for events with $M(pK^-)>$ 1.55
GeV/$c^{2}$. A fit to the RMM background
spectrum only (dashed line) and with a Gaussian function (solid
line) in the region below 1.65 GeV/$c^{2}$. The dotted line is 
the background.  
(b) The background spectra for the best fits to RMM spectra 
with the wide signal region (dashed line), the narrow region (dotted line),
and the default region (solid line). 
}
\label{fig:varseg}
\end{figure}

Fig.~\ref{fig:nk}(b) shows the Dalitz plot of $M^2(nK^+)$ \textit{vs.}
$M^2(pK^-)$. Note a proton is assumed to be a spectator for the
calculation of $M(nK^+)$ and a neutron is assumed to be a spectator
for the calculation of $M(pK^-)$. The relatively large -2ln$L$ values for
the fits of the $M(nK^+)$ distribution could be due to the reflections
of the $\Lambda $(1520) events which might be responsible for the
remaining structure near 1.65 GeV/$c^{2}$. To avoid a possible effect
due to the reflection, we require events to satisfy $M(pK^-)>$ 1.55
GeV/$c^{2}$ and restrict the fit region up to 1.65 GeV/$c^{2}$. The
fit qualities are improved, giving -2ln$L$/$ndf$=55.2/33 and
-2ln$L$/$ndf$=24.8/31 for the cases with and without the
$\Theta ^{+}$ contribution, respectively. The significance is
unchanged because the change in $\Delta$(-2ln$L$) is small.
Fig.~\ref{fig:varseg}(a) shows the $M(nK^+)$ distribution after
the $\Lambda (1520)$ exclusion cut.

To study the model dependence, we have varied the
boundaries of the seed regions for the RMM spectrum generation: the
narrow signal region case with the boundaries at 1.51 GeV/$c^{2}$ and
1.54 GeV/$c^{2}$, and the wide signal region case with the boundaries
at 1.48 GeV/$c^{2}$ and 1.57 GeV/$c^{2}$. The fit results are
essentially unchanged, giving a statistical significance of 5.2
$\sigma $ for the narrow signal region case, and a significance of 5.1
$\sigma $ for the wide signal region case.  Although the shape
and magnitude of each RMM spectrum vary case by case, the resultant
summed background spectrum for the best fit is similar to each other
as shown in Fig.~\ref{fig:varseg}(b). Fine structures in the original
$M(nK^+)$ distribution compared to a typical mass
spread of $\sim $30 MeV/$c^{2}$ due to Fermi motion are smoothed by
using the RMM. The weak dependence of the fit results on the seed
boundary condition is a consequence of the smooth nature of the
uncorrelated background. The maximum difference in the fitted peak 
height with the various background models is approximately 5 $\%$, 
which is much smaller than the fitting uncertainty of 18 $\%$, and,
therefore, neglected. 

\begin{figure}[htbp]
\begin{tabular}{c c}
\begin{minipage}{0.5\hsize}
\begin{center}
 \includegraphics[width=7.5cm,height=7.5cm,keepaspectratio]
{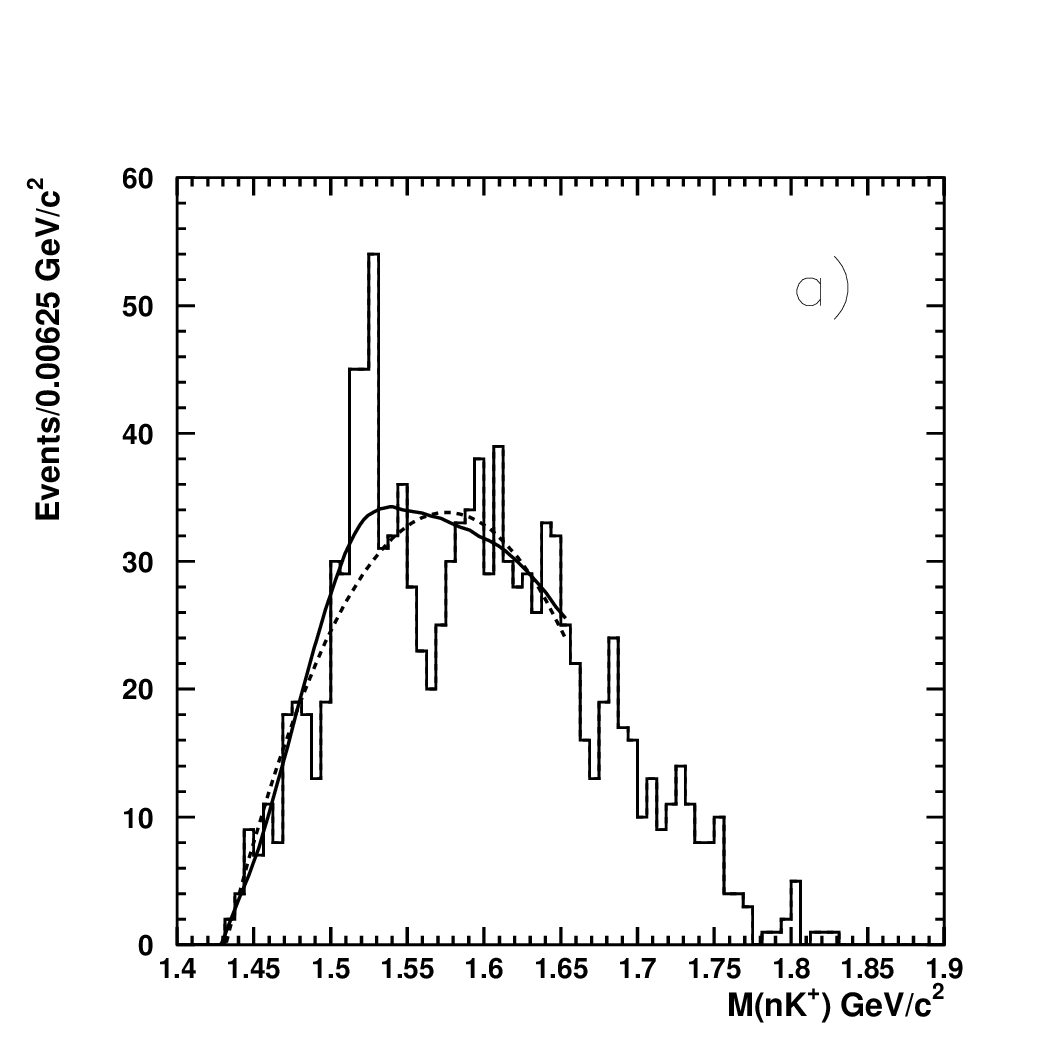}
\end{center}
\end{minipage} & %
\begin{minipage}{0.5\hsize}
\begin{center}
 \includegraphics[width=7.5cm,height=7.5cm,keepaspectratio]
{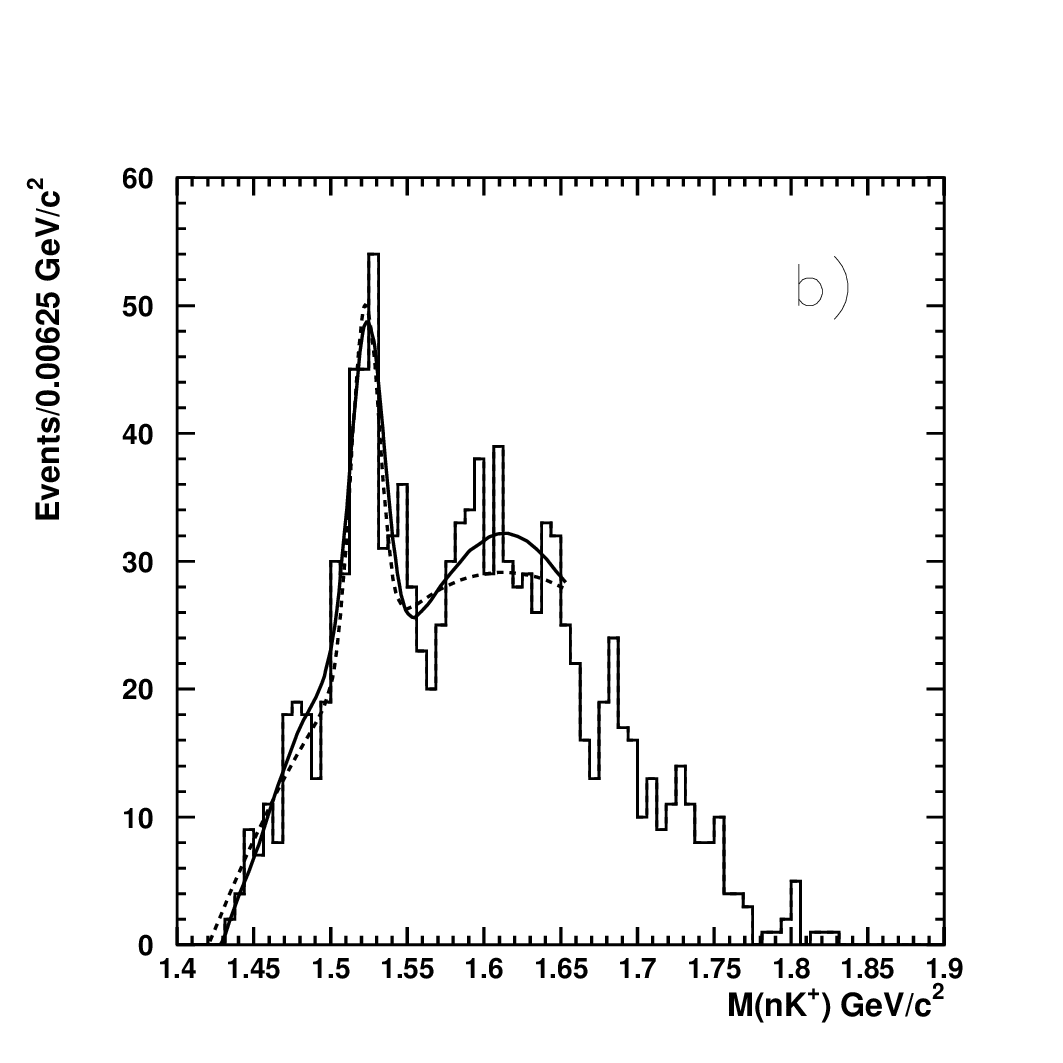}
\end{center}
\end{minipage} \\
\begin{minipage}{0.5\hsize}
\begin{center}
 \includegraphics[width=7.5cm,height=7.5cm,keepaspectratio]
{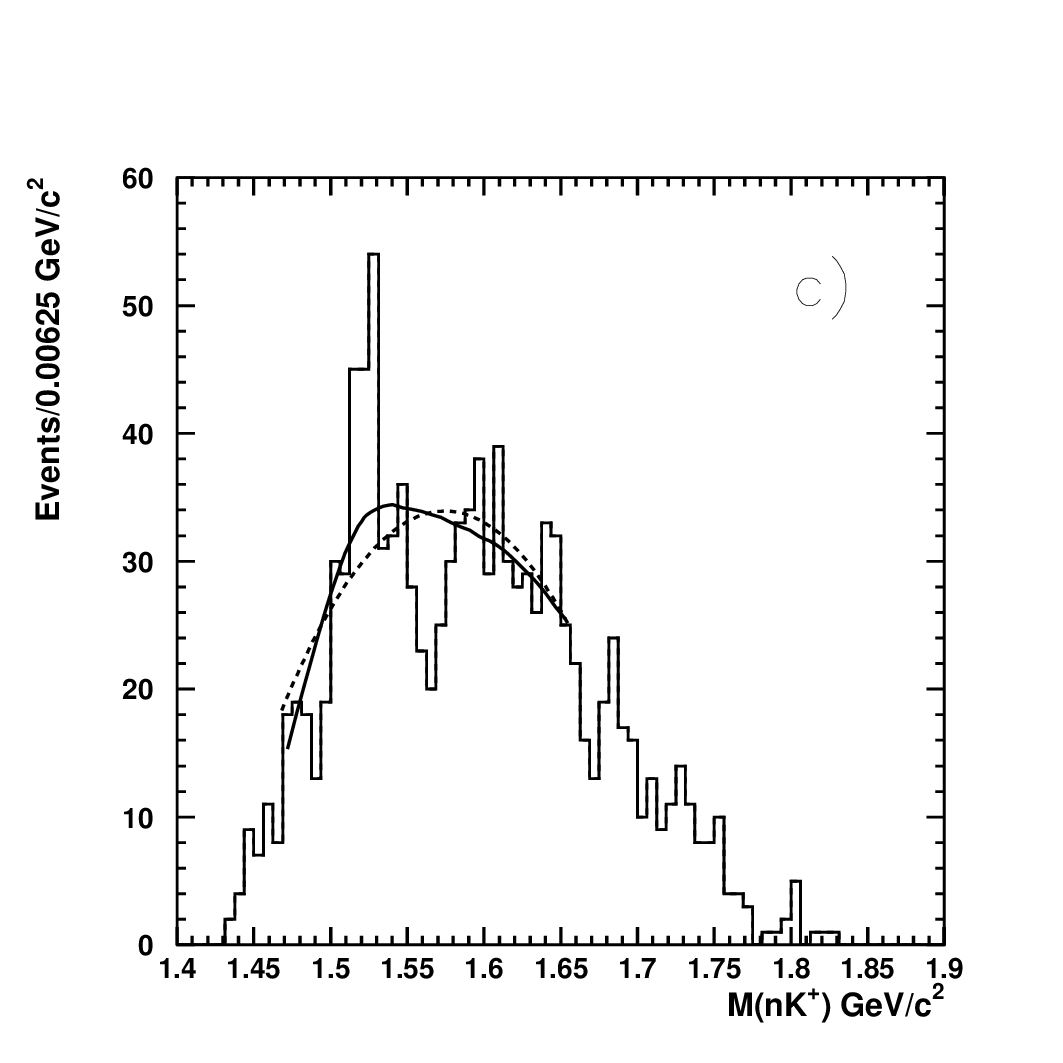}
\end{center}
\end{minipage} & %
\begin{minipage}{0.5\hsize}
\begin{center}
 \includegraphics[width=7.5cm,height=7.5cm,keepaspectratio]
{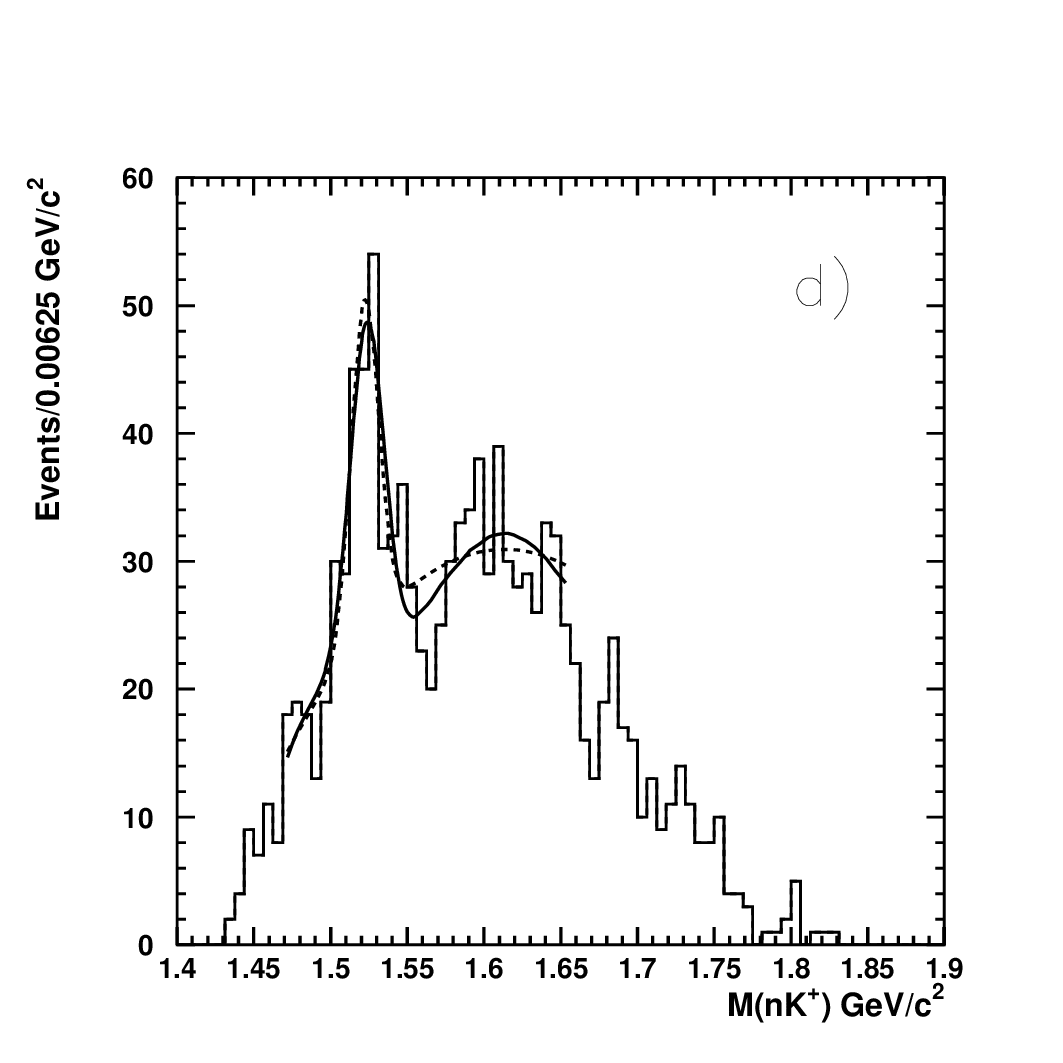}
\end{center}
\end{minipage} \\
\end{tabular}
\caption{ Comparison of the fits with the RMM distributions (solid line)
and a 2nd-order polynomial functions (dashed line): (a) in the region of 1.43
GeV/$c^{2} <M(nK^+)<$ 1.65 GeV/$c^{2}$ without the $\Theta^+$
contribution. (b) with the $\Theta^+$ contribution. (c) in the region
of 1.47 GeV/$c^{2} <M(nK^+)<$ 1.65 GeV/$c^{2}$ without the $\Theta^+$
contribution. (d) with the $\Theta^+$ contribution.  }
\label{fig:polfit}
\end{figure}

We have also examined a fit to the mass distribution using a
second-order polynomial to represent the background. Note the number
of the fitting parameters for the fit is the same as the fit with
three RMM background spectra. Fig.~\ref{fig:polfit} shows the
comparison of the fit results for the cases with the fitting regions
of 1.43 GeV/$c^{2} <M(nK^+)<$ 1.65 GeV/$c^{2}$ and 1.47
GeV/c$^{2} <M(nK^+)<$1.65 GeV/c$^{2}$. Fit quality is always
better with the RMM than with the polynomial background function. By
using the polynomial background function, the -2ln$L$/\textit{ndf
}values for the wide fitting region are obtained to be 65.1/33 and
28.1/31 without and with the $\Theta ^{+}$ contribution,
respectively. The $\Delta $(-2ln$L)$ of 37.0 corresponds to a
5.7$\sigma $ significance. For the narrow fitting region case, the
-2ln$L$/\textit{ndf }values for the fits without and with the $\Theta^+$ 
contribution are 58.4/27 and 23.1/25, resulting
in a 5.6 $\sigma $ significance. For the same fitting region, the
-2ln$L$/\textit{ndf } values using the RMM are 51.2/27 and 21.2/25,
giving a significance of 5.1 $\sigma$. Thus, the statistical
significances estimated from the fit results with RMM are smaller than
those with the polynomial functions. The difference is caused by poor
modeling of a background shape with the polynomial functions,
especially without the $\Theta ^{+}$ contribution.
The fit results with the various background models are summarized 
in Table~\ref{fitsummary}. The smallest significance of 5.1 $\sigma$
is considered as the $\Theta^+$ significance with the systematics 
taken into account.

{\small
\begin{table}[tbp]
\caption{Summary of fit results that are used to study the significance of 
the $\Theta^+$ contribution.}
\label{fitsummary}
\begin{tabular}{l l r r r r}
\hline
\hline
Background model & $\Lambda(1520)$ cont. & Fit region & -2ln$L$/$ndf$  & -2ln$L$/$ndf$ & Significance \\
      & &  (GeV/$c^{2}$) & without $\Theta^+$ & with $\Theta^+$ &  \\
\hline
RMM spectra, default seed sets. & not excluded & [1.43,1.85] & 
104.7/66 & 73.64/64 & 5.2 $\sigma$ \\
RMM spectra, default seed sets. & excluded & $[1.43,1.65]$ & 
55.2/33 & 24.8/31 & 5.2 $\sigma$ \\
RMM spectra, wide signal region. & excluded & $[1.43,1.65]$ & 
54.5/33 & 24.3/31 & 5.1 $\sigma$ \\
RMM spectra, narrow signal region. & excluded & $[1.43,1.65]$ & 
55.9/33 & 24.8/31 & 5.2 $\sigma$ \\
RMM spectra, default seed sets. & excluded & $[1.47,1.65]$ & 
51.2/27 & 21.2/25 & 5.1 $\sigma$ \\
2nd-order polynomial. & excluded & $[1.43,1.65]$ & 
65.1/33 & 28.1/31 & 5.7 $\sigma$ \\
2nd-order polynomial. & excluded & $[1.47,1.65]$ & 
58.4/27 & 23.1/25 & 5.6 $\sigma$ \\
\hline
\hline
\end{tabular}
\end{table}
}
  
The validity of the statistical significance estimated from the $\Delta 
$(-2ln$L)$ value is checked with $2\times 10^6$ sample mass distributions 
generated by a toy Monte-Carlo simulation program by assuming a spectrum shape
for the non-resonant $K^+K^-$ events. The generated distributions are fitted to the mass 
distribution which uses the polynomial background function. The difference 
of the -2ln$L$ values with and without a Gaussian function with the 
width of 11 MeV/$c^{2}$ is checked. The numbers of the samples with a 
significance of more than 4$\sigma $ and 5$\sigma $ are 10 and 2, 
respectively. These numbers are consistent with the expected number of 
occurrences of high-significance samples due to statistical fluctuations. 

\begin{figure}[htbp]
\begin{tabular}{c c}
\begin{minipage}{0.5\hsize}
\begin{center}
 \includegraphics[width=7.5cm,height=7.5cm,keepaspectratio]
{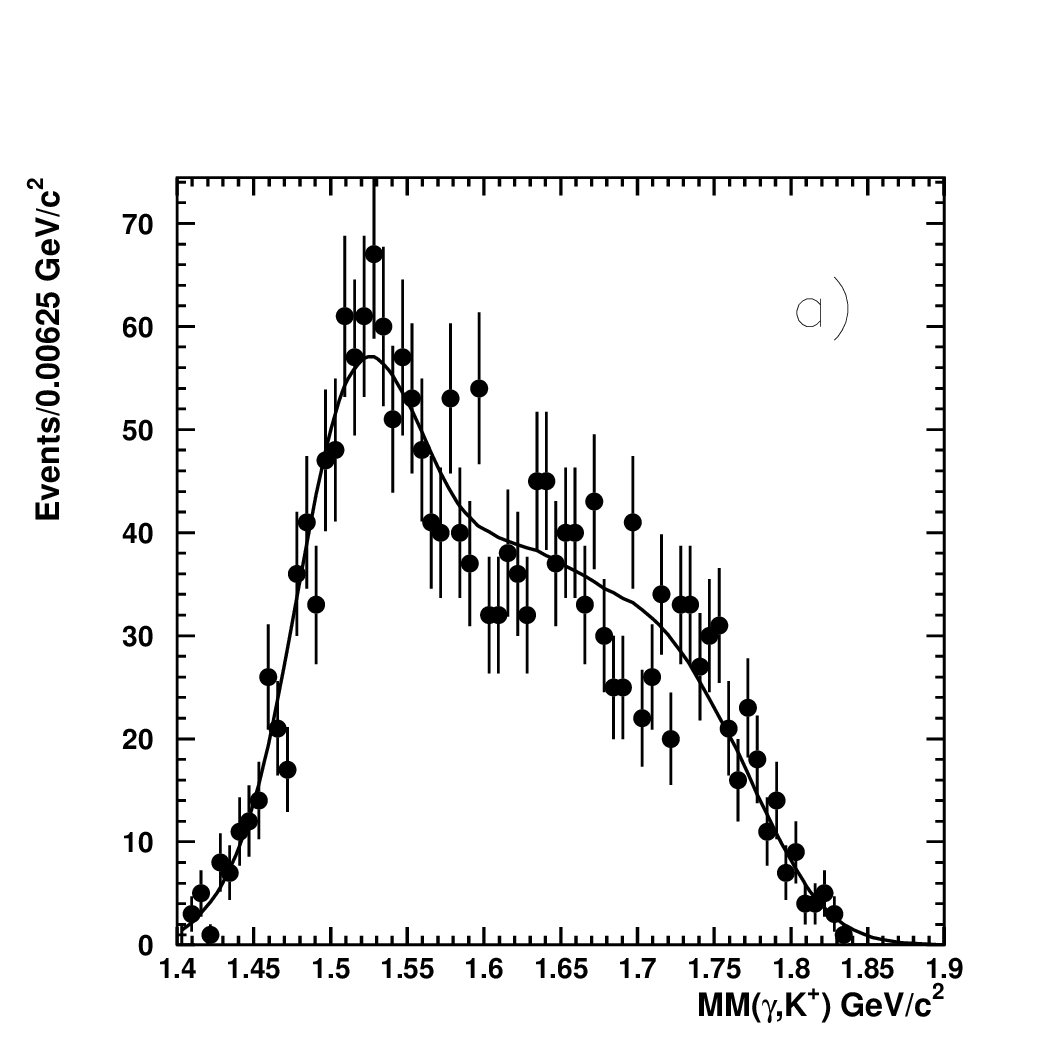}
\end{center}
\end{minipage} & %
\begin{minipage}{0.5\hsize}
\begin{center}
 \includegraphics[width=7.5cm,height=7.5cm,keepaspectratio]
{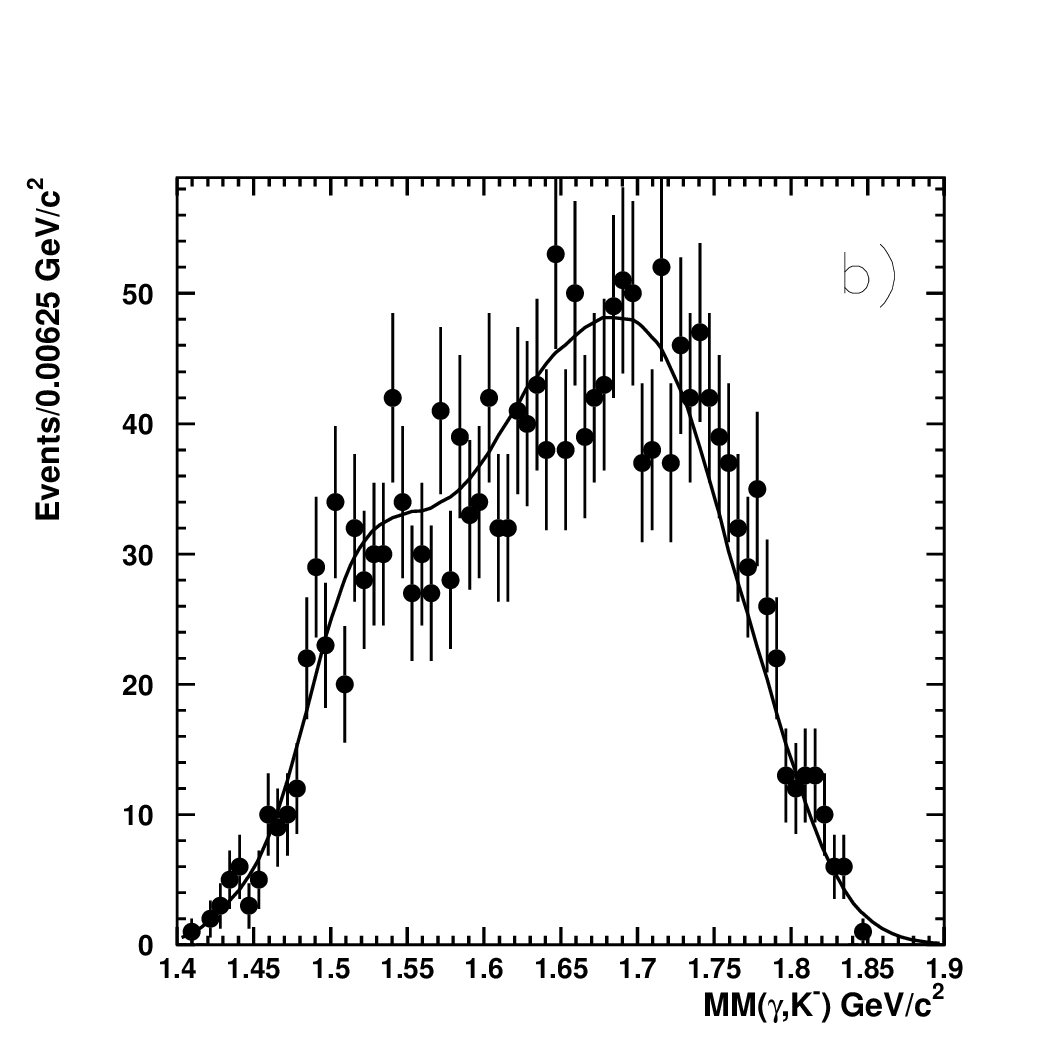}
\end{center}
\end{minipage}
\end{tabular}
\caption{$MM(\gamma ,K^+)$ (left) and $MM(\gamma ,K^-)$ (right) distributions
with a fit to a mass distribution consisting of reversed RMM spectra (solid line).
}
\label{fig:mmfna}
\end{figure}

A background spectrum for $MM(\gamma ,K^-)$ and $MM(\gamma
,K^+)$ distributions can be obtained by using the RMM in a reversed
way, where a measured $M(nK^+)$ ($M(pK^-))$ and a randomized $p_{\min
}$ are used to simulate $MM(\gamma ,K^-)$ ($MM(\gamma ,K^+))$.  
The missing mass distributions with a fit to the
reversed RMM functions are shown in Fig.~\ref{fig:mmfna}. Both
$MM(\gamma ,K^+)$ and $MM(\gamma ,K^-)$ distributions are well reproduced 
by the background functions with -2ln$L$/$ndf$=79.7/67 and 60.2/66,
respectively.  Since the -2ln$L$ values are not reduced by more than
$\Delta ndf$ by including a Gaussian function with a fixed
width of 30 MeV/$c^{2}$, the corresponding significances are less than
1. This demonstrates the importance of the narrowness of the width and
the consistency between the measured and estimated values for the
shape analysis. 

\begin{figure}[htbp]
\begin{tabular}{c c}
\begin{minipage}{0.5\hsize}
\begin{center}
 \includegraphics[width=7.5cm,height=7.5cm,keepaspectratio]
{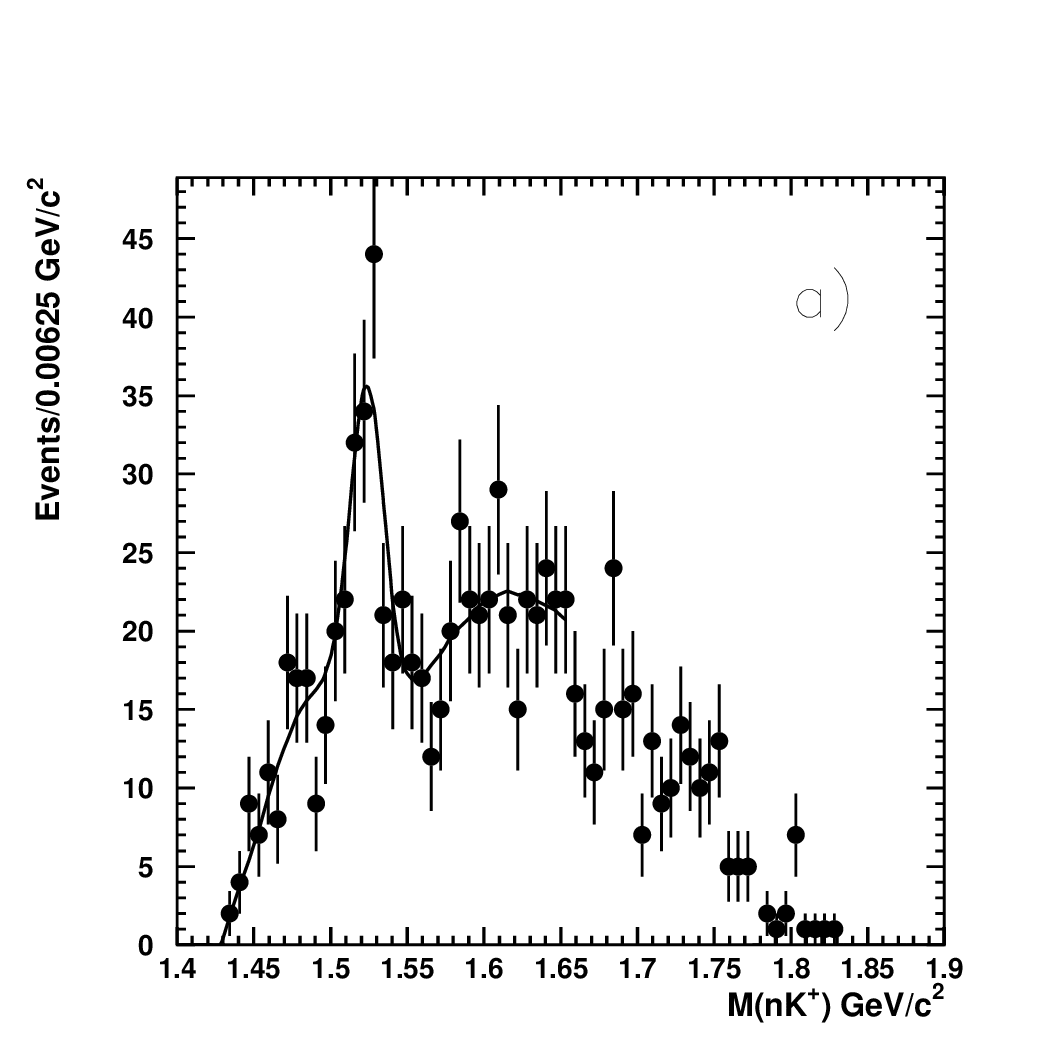}
\end{center}
\end{minipage} & %
\begin{minipage}{0.5\hsize}
\begin{center}
 \includegraphics[width=7.5cm,height=7.5cm,keepaspectratio]
{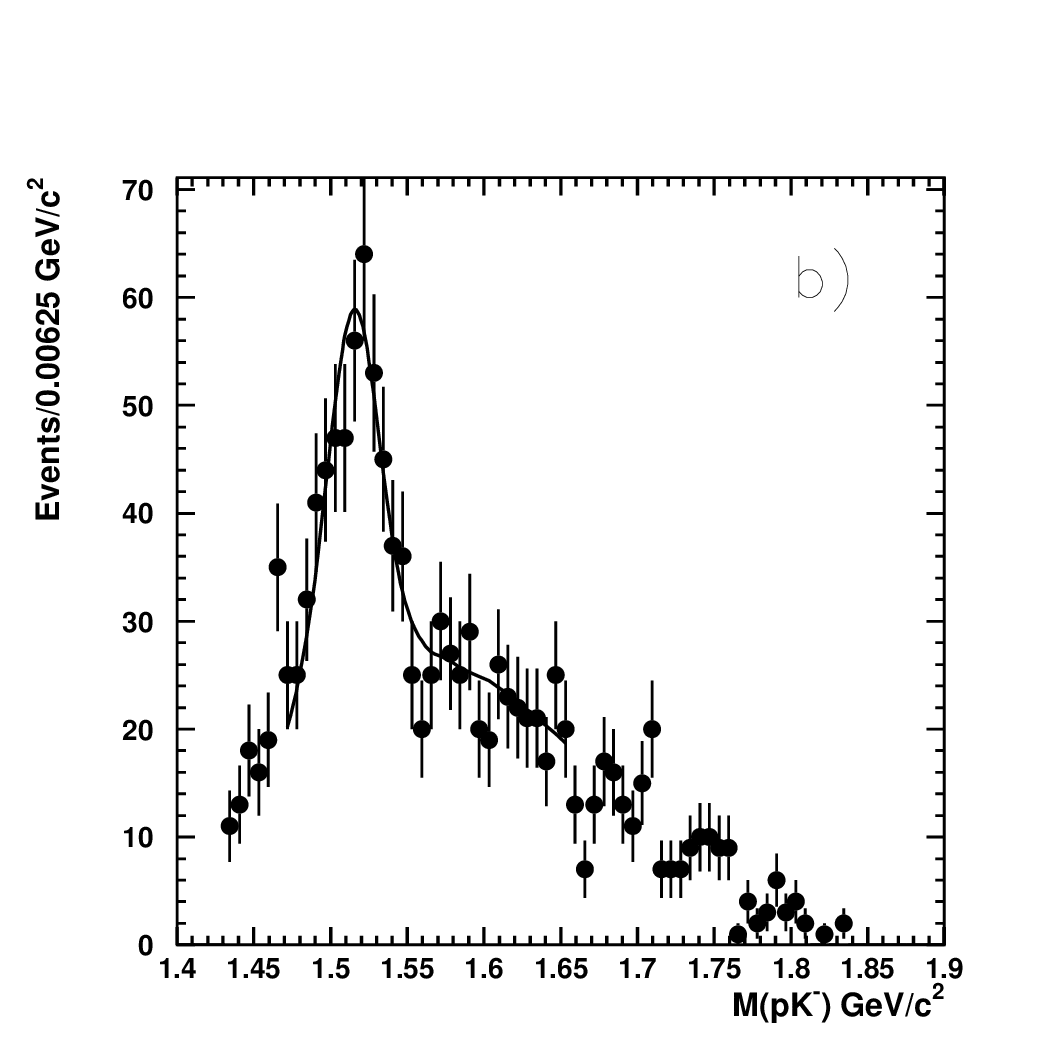}
\end{center}
\end{minipage}
\end{tabular}
\caption{
$M(nK^+)$ (left) and $M(pK^-)$ (right) distributions for events with
$M(K^+K^-)>$1.05 GeV/$c^{2}$. The solid lines are fits to the RMM functions
plus a Gaussian function. 
}
\label{fig:kk105}
\end{figure}

A photon energy independent $\phi $ exclusion cut with the condition of
$M(K^+K^-)>$1.05 GeV/$c^{2}$ is also tried, and the resultant $M(nK^+)$ and
$M(pK^-)$ distributions are shown in Fig.~\ref{fig:kk105}. Both
distributions are well fitted to the mass distributions with the RMM
and Gaussian functions with -2ln$L$/$ndf$ ratios of less than
1. The peak positions have not been changed, and the signal yields for
the $\Theta ^{+}$ and $\Lambda $(1520) are reduced by a factor of
25{\%} and 35{\%}, respectively, compared to those obtained by the
original $\phi $ exclusion cut.

Events with a 3rd charged track are examined to check if the narrow peak in 
the $M(nK^+)$ distribution is due to quasi-free reactions with a spectator 
proton. The ratio of the number of events with a 3rd track to the total 
number of $K^+K^-$ events is 2.9$\pm $0.9 {\%} in the mass region of 1.50 
GeV/$c^{2} <M(nK^+)<$ 1.55 GeV/$c^{2}$, while it is 8.8$\pm $1.2{\%} in the 
mass region of 1.50 GeV/$c^{2} <M(pK^-)<$ 1.55 GeV/$c^{2}$, where the 
$\Lambda $(1520) events dominate. The 3rd track ratios in the adjacent 
regions of 1.55 GeV/$c^{2} <M(nK^+)<$ 1.60 GeV/$c^{2}$ and 1.55 
GeV/$c^{2} <M(pK^-)<$ 1.60 GeV/$c^{2}$ are 7.2$\pm $1.6 {\%} and 6.4$\pm 
$1.4 {\%}, respectively.
A fit to the $M(nK^+)$ distribution for events without the 3rd track 
exclusion shows that the $\Theta^+$ peak height varies only by +0.9 {\%}, while the 
background level increases by +4.6{\%}. The $\Lambda $(1520) peak height 
decreases by 8.9 {\%} by removing events with a 3rd track. 
These observations indicate the $\Theta^+$ peak is likely due 
to quasi-free $\gamma n$ reactions.

In our previous paper, the statistical significance has been estimated
from the ratio of $S/\sqrt B $, where $S$ and $B$ are the numbers of
the signal and background events, respectively~\cite{LEPS}. The toy
Monte-Carlo study has shown this method results in large
overestimation of the significance. The magnitude of the
overestimation is still large when using $S/\sqrt {S+B} $ instead of
$S/\sqrt B $. In the current analysis, the $S/\sqrt {S+B}$ value in the
mass region of 1.50 GeV/$c^{2}$ $<M(pK^-)<$ 1.55 GeV/$c^{2}$ is 6.9,
which is larger than the significance estimated from the $\Delta
$(-2ln$L$) by approximately 2. The ratio of the peak height to its
fitting uncertainty gives a significance of 5.4 $\sigma$ which is slightly
($\sim $5{\%}) higher than that estimated from $\Delta $(-2ln$L$).

\section{CONCLUSIONS AND DISCUSSIONS}
We have observed a narrow peak near 1.53 GeV/c$^{2}$ in the $nK^+$ invariant 
mass distribution from quasi-free $\gamma n\to K^+K^-n$ reactions. The 
Fermi-motion corrected mass distribution is obtained by using the newly 
developed minimum momentum spectator approximation (MMSA). The validity of 
the MMSA is checked by analyzing the quasi-free $\gamma p\to K^+\Lambda 
(1520)$ reactions. The effect of the Fermi-motion on the \textit{nK}$^{+}$ invariant 
mass is studied by using the randomized minimum momentum method (RMM), and 
it has been shown a narrow peak with a width much less than $\sim$ 30 MeV/$c^{2 
}$ cannot be generated by corrections nor selection cuts. The statistical 
significance of the $\Theta ^{+}$ peak has been estimated by a spectrum 
shape analysis using the RMM background functions as well as polynomial 
functions. The statistical significance from the shape analysis is 5.1 
$\sigma$. 

The $\Theta^+$ yield is estimated to be $116 \pm 21$ events in the
$1.5 \times 10^4$ $K^+K^-$ events.  The differential cross-section is
estimated to be $12 \pm 2$ nb/sr in the LEPS angular range by
assuming the isotropic production of the $\Theta^+$ in the $\gamma n$
center-of-mass system.

The $\Theta^+/K^+K^-$ ratio of $(0.8 \pm 0.1) \times 10^{-2}$ is
consistent with that of $(1.1 \pm 0.2) \times 10^{-2}$ obtained by the
previous measurement although the detector acceptance is smaller in
the current experiment due to a longer distance from the target to the
spectrometer~\cite{LEPS}.  However, the significance of the $\Theta^+$
contributions in the previous study is highly overestimated because it
was calculated from the $S/\sqrt{B}$ ratio.

The yield ratio of the $\Theta ^{+}$ to the $\Lambda $(1520) is
0.40$\pm $0.09. By considering the partial decay branching ratios of
$\Gamma (\Lambda (1520)\to N\overline K )/\Gamma (\Lambda (1520)\to
all)=0.45$ and the acceptance difference, the production ratio of the
$\Theta ^{+}$ to the $\Lambda $(1520) is estimated to be 0.15$\pm
$0.03 in our detector acceptance.

The upper limit on the $\Theta ^{+}$ production cross-section obtained
by the CLAS collaboration is 3 nb~\cite{McKinnon2006}.  However,
due to the different nature of the measurements at CLAS and LEPS, the
CLAS upper limit is very difficult to compare with the present
results, since the re-scattering mechanism of the spectator proton
(required by CLAS) is unknown and the detector acceptances are almost
exclusive.  Hence we see no conflict between the present results and
those published by CLAS.

The LEPS collaboration will analyze new data which were collected with the 
same detector setup and an improved luminosity (by a factor of 3). Detailed 
investigation of the angular and energy dependencies of the $\Theta ^{+}$ 
photo-production will become possible if the peak is confirmed in the new 
data set. 

\section{ACKNOWLEDGEMENT}
We thank the staff at SPring-8 for providing a stable beam and
excellent experimental conditions.  We thank Dr.~A.~Hosaka and
Dr.~A.~I.~Titov for helpful discussions.  This research was supported
in part by the Ministry of Education, Science, Sports and Culture of
Japan, by the National Science Council of the Republic of China
(Taiwan), by the National Science Foundation (USA), and by the Korea
Research Foundation (Korea).

\end{document}